# Naturally Occurring Genetic Variation Influences the Severity of Drosophila Eye Degeneration Induced by Expression of a Mutant Human Insulin Gene

Sarah Carl[1, 2]

[1]Department of Ecology and Evolution, The University of Chicago, Chicago, IL, 60637
[2]Current affiliation: Department of Genetics, The University of Cambridge, Cambridge, UK, CB2 3EH
E-mail: s.carl@gen.cam.ac.uk

This thesis was submitted to the Undergraduate Honors program in Biology at the University of Chicago in June, 2010. Many people contributed to this work on an intellectual and technical level. The ideas and experimental design described herein originated from discussions between Martin Kreitman and Graeme Bell. All transgenic lines of *Drosophila* were created by Michael Ludwig and Soo-Young Park. Michael Ludwig helped develop the crossing scheme and played a primary supervisory role in the laboratory, assisting with experimental protocols and *Drosophila* husbandry. My gratitude and respect is due to all of them for their support during the experiments and the writing process.

**Abstract**

Dominant negative mutations in the insulin gene are the second most common cause of permanent neonatal diabetes. However, variation in severity and penetrance of neonatal diabetes, as in other complex genetic diseases, cannot be accounted for by known "disease" mutations. In a novel approach to this problem, we have utilized the genetic tools available in *Drosophila* to model the effects of the C96Y mutation, a cysteine to tyrosine mutation in the insulin protein that can cause permanent neonatal diabetes in humans. This mutation, which disrupts a disulfide bridge in the proinsulin molecule, has been shown to lead to partial protein unfolding and aggregation in the endoplasmic reticulum. It is thought to induce β-cell death in humans and mice through endoplasmic reticulum stress-mediated apoptosis. We employed the *UAS/GAL4* system to create a stable fly laboratory stock expressing human insulin with the C96Y mutation (*INS$^{C96Y}$*) in the *Drosophila* eye. We crossed these transgenic *INS$^{C96Y}$* flies to 36 isogenic lines derived from a wild population. The F1 flies displayed a disrupted eye development phenotype, with both reduced eye size and irregularity in ommatidia, which varied between lines and with sex. By backcrossing F1 offspring to their wild parents, we were able to analyze the effects of naturally occurring genetic variation on the *INS$^{C96Y}$* phenotype in flies both heterozygous and homozygous for wild third chromosomes. We hypothesized that the physiological stress placed on eye cells expressing *INS$^{C96Y}$* would reveal the effect of background genetic variation, visible as variation in phenotype. We measured four different characters in *INS$^{C96Y}$* eyes to quantify the extent of disruption of development: the size of the eye, as standardized against thorax length, the shape of the eye, the presence or absence of lesions, and the degree of ommatidial structure present. We observed significant differences between lines in terms of all four of these characters, indicating that an approach that takes into account quantitative variation in the genetic background can yield valuable insights into the phenotypic severity and penetrance of a complex genetic disease.

*Keywords: Drosophila,* insulin, proinsulin, diabetes, buffering, ER stress

**Introduction**

Complex genetic diseases show a wide range of variation in severity and penetrance in human populations; while genetic factors clearly play an important role in generating this variation, attempts to identify responsible loci have been unsuccessful for the majority of diseases with a genetic component (McClellan and King, 2010). With the exception of diseases that show patterns of simple Mendelian inheritance, those loci that have been identified, particularly through genome-wide association studies, tend to have a weak ability to predict disease (McClellan and King, 2010). Loci that have been identified in association with complex genetic diseases generally have only a small effect on the phenotype of those diseases and represent only a small portion of the diseases' heritable component (Kruglyak, 2008). This is true for diabetes mellitus. Previous studies attempting to characterize the genetic component of diabetes have failed to address the interaction of loci known to be associated with diabetes with quantitative variation in the genetic backgrounds of individuals. Such interaction can significantly contribute to the complexity of genetic diseases, as well as cause population-restricted effects, as different populations may exhibit differing frequencies of certain polymorphisms. Indeed, failure to account for population stratification may be one major limiting factor in previous approaches to identifying the genetic architecture of complex diseases (McClellan and King, 2010). We hypothesized that, by examining the effect of naturally occurring background genetic variation on the diabetes phenotype in a model organism, we would be more successful in recovering polymorphisms that, while not in themselves "disease" genes, contribute in a piecewise manner to the enhancement or suppression of the disease phenotype.

Diabetes mellitus is a group of metabolic diseases with diverse causes that is most broadly characterized by high blood glucose levels. Many forms of diabetes have important genetic components, though monogenic forms of the disease are relatively rare (Støy, et al., 2007). Neonatal diabetes is a rare form of diabetes, occurring once in every 215,000 – 500,000 live births. It is typically diagnosed within the first 6 months of life (Støy, et al., 2007; Edghill, et al., 2008). Though it presents with similar symptoms as the more common type 1 diabetes, including hyperglycemia and diabetic ketoacidosis, neonatal diabetes is not associated with autoantibodies for the pancreatic β-cells, as is type 1 diabetes (Edghill, et al., 2008). The most common cause of neonatal diabetes is a mutation in the gene that codes for one of the two protein subunits of the ATP-sensitive potassium channel of the β-cell (Støy, et al., 2007). However, several studies have recently shown that it can also be caused by dominant mutations in the insulin gene itself (Støy, et al., 2007; Edghill, et al., 2008). The majority of mutations in the insulin gene that have been recovered in neonatal diabetes patients occur in regions of the preproinsulin molecule that may prevent its proper folding and processing through the insulin secretory pathway (Støy, et al., 2007). We chose to examine one of these mutations, the C96Y mutation, which is a cysteine to tyrosine substitution at residue 96 in the preproinsulin molecule.

Insulin is biosynthesized in pancreatic β-cells via a series of steps mediated by the endoplasmic reticulum (ER). A single-chain polypeptide, preproinsulin, is encoded by the insulin gene. This molecule is then translocated to the ER, where its signal peptide is cleaved to yield proinsulin. Proinsulin is folded through an ER-mediated oxidative process, during which the specific pairing of three disulfide bridges is a crucial step. The three disulfide bridges link together the two polypeptide chains of mature insulin after the connecting C-peptide is cleaved and are necessary for its stability and bioactivity (Weiss, 2009). Replacing the cysteine at position 96 with a tyrosine disrupts the formation of one of the disulfide bridges linking the A chain and the B chain (Fig. 1). In the Akita mouse model, in which the same mutation has been observed in the mouse *Ins2* gene, the C96Y mutation leads to partial protein unfolding and aggregation in the ER, in turn causing defects in proinsulin trafficking, increased ER stress, mitochondrial swelling, and eventual β-cell death (Weiss, 2009). In vitro studies expressing the human insulin gene with the C96Y mutation (*INS$^{C96Y}$*) in MIN6 mouse insulinoma cells and rat INS-1 insulinoma cells have demonstrated that the mutant insulin is completely retained in the ER and fails to exit into secretory granules or undergo proteolytic processing. A dominant-negative effect of *INS$^{C96Y}$* on the production and secretion of wild-type insulin has also been observed. In addition, cells expressing *INS$^{C96Y}$* display increased splicing of XBP1 as well as elevated levels of ER chaperones such as Grp78 and the protein Chop, all of which are indicators of ER stress (Park, et al., 2010; Rajan, et al., 2010; Colombo, et al., 2008). All of these results indicate that the disruption of proper proinsulin folding caused by the C96Y mutation compromises insulin secretion and general β-cell function. Although the mechanism of β-cell failure and resulting permanent neonatal diabetes due to the C96Y mutation has not been extensively characterized in humans, it is believed to follow a very similar pathogenesis (Weiss, 2009). The C96Y mutation thus represents a well-described genetic factor involved in generating neonatal diabetes, which lends itself to further analysis of the effects of genetic variation on the manifestation of the disease.

Here we approach the question of modeling diabetes in a novel way through utilizing the genetic tools available in *Drosophila melanogaster*. We have created a line of *Drosophila* expressing *INS$^{C96Y}$* under an eye-specific driver, *GMR*, using the *UAS/GAL4* system on the second chromosome. These *INS$^{C96Y}$* flies display disruption in eye development, visible as both a reduction in overall size and a degeneration of ommatidial structure. We expected that the expression of $INS^{C96Y}$ would have a similar effect on *Drosophila* eye cells as it has been shown to have on β-cells and cultured insulinoma cells, placing stress on the protein folding and trafficking machinery of the cells, in particular the ER. In situations of extreme ER stress, when the amount of unfolded or aggregated protein present exceeds the ER's capacity for processing or degradation, the unfolded protein response can induce cell death by apoptosis (Yoshida, 2007; Nozaki, et al., 2004). The basic mechanism of the unfolded protein response is conserved among all eukaryotes, making it relevant to the study of the $INS^{C96Y}$ phenotype in both flies and mammals. ER stress and resulting cell death at various points during the developmental process is one mechanism that could result in the observed phenotype of eye degeneration in $INS^{C96Y}$ flies. We therefore expected that variation in genes involved in protein folding and the ER stress response would be particularly manifest as variation in phenotypic severity.

    The fly eye is a particularly well-suited system for observing the effects of unfolded protein on development. It has a highly stereotyped developmental program, which has been thoroughly characterized. The wild-type fly eye develops into a highly regular array of ommatidia, each with eight photoreceptor neurons. This program is generally considered to be a strong example of developmental buffering; many well-known mutations that reduce eye size or pigmentation still result in an eye with regular, differentiated ommatidia. The introduction of mutant insulin disrupts this pattern, resulting in an easily visible phenotype. A similar system has been used to model neurodegenerative diseases caused by polyglutamine expansions in ataxin proteins, which

also cause misfolding and protein instability (Bilen and Bonini, 2007; Lessing and Bonini, 2008). After observing a disrupted eye development phenotype in the transgenic $INS^{C96Y}$ flies, we asked if we could reveal quantitative differences in the severity of that phenotype by placing it in different wild genetic backgrounds, which represented a sample of the total quantitative genetic variation present in the population. We hypothesized that the physiological stress placed upon the cells of the *Drosophila* eye by the presence of mutant insulin would sensitize them to natural genetic variation, overcoming the buffering of the developmental process and leading to visible, quantitative phenotypic variation. In order to test this hypothesis, we crossed the $INS^{C96Y}$ stock flies to 36 inbred, isogenic lines of *Drosophila melanogaster* collected from a single wild population in North Carolina (Jordan, et al., 2008; Mackay, et al., 2012). We then backcrossed F1 offspring to their wild parents, allowing us to analyze variation in the severity of eye degeneration in F2 flies both heterozygous and homozygous for a wild third chromosome (See Fig. 2 for crossing scheme).

The genetic interactions and regulatory relationships required to produce a complex trait are highly intricate. Rather than identifying a "disease" gene, we have examined how variation in input from a wide range of genes in the genetic background can result in quantitative, visible variation in the output of the disease phenotype. Our approach also highlights the importance of genetic diversity, including potentially rare variants, within a population. On an evolutionary level, phenotypes with differential fitnesses are the raw material upon which natural selection can act. Our demonstration that variation in the genetic background can produce significant differences in the severity of a disease phenotype caused by a particular allele supports the idea that such quantitative genetic variation can be important in generating adaptations as well as mitigating the effect of a deleterious mutation. As genetic testing for diseases and personalized medicine based on an individual's genetic sequence become more commonplace, it will be important to take into account the complexity of the genetic interactions that go into producing a disease. Such considerations will allow medical professionals to design more personalized, effective treatments and screens for degree of risk. Here we report quantitative differences uncovered in the severity of the $INS^{C96Y}$ phenotype in different genetic backgrounds, all drawn from one wild population. We anticipate that, through further investigation of polymorphisms present in these wild lines, new insights will be gained into the origin and possible treatment of permanent neonatal diabetes.

## Materials and Methods

*Fly Stocks*

The $INS^{C96Y}$ stock was created by Michael Ludwig at the University of Chicago. Two transgenic lines were obtained from fly stocks: one line carrying a *GMR-GAL4* driver *P*-element on the second chromosome and one line carrying a *UAS-GFP* responder *P*-element on the second chromosome. Both *P*-elements contained a $w_+$ dominant marker. A third line was engineered by the Bell and Kreitman labs that carried a *UAS-INS*$^{C96Y}$ responder *P*-element on the second chromosome, also with a $w_+$ marker. All three lines were generated from the *w1118* laboratory line, which has a $w_-$ background. These lines were crossed to generate a recombinant line with all three transgenes on the same second chromosome, which was balanced over the *CyO* balancer chromosome. Wild lines, numbered 174 through 209, 445, 744, and 745, were obtained from Trudy Mackay at North Carolina State University in Raleigh, NC. These lines, which represent a subset of the Drosophila Genetic Reference Panel (DGRP) were derived from flies originally collected from the Raleigh, NC Farmer's Market in 2003. Inbred lines were established through 20 generations of full-sib inbreeding (Jordan et al., 2007; Mackay, et al., 2012). All stocks were maintained at 25°C on standard cornmeal medium.

*Crosses*

A detailed crossing scheme is represented in Figure 2. For the first cross, 4-6 males from each wild line were collected and placed into two duplicate vials of medium per line, for a total of 2-3 males per vial. These males were mated with 5 virgin females from the $INS^{C96Y}$ stock per vial. The parent flies were transferred into new vials after 2 days. Offspring were visually inspected, and those without the $INS^{C96Y}$ transgene, which were marked by the *CyO* balancer chromosome, were discarded. Of F1 offspring with the $INS^{C96Y}$ transgene, flies from four classes were collected: males with the third chromosome balancer *TM3, Sb* (marked by the *Sb* mutation, which produces stubbled bristles), males without *TM3, Sb*, females with *TM3, Sb*, and females without *TM3, Sb*. A goal was set of 5 flies per class per line, though in some lines it was impossible to collect 5 from each class due to decreased viability of flies with the *TM3, Sb* balancer. These lines were still included in analyses, but with a smaller number of individuals. Collected flies were placed immediately into a freezer at -80°C. Males with both the $INS^{C96Y}$ transgene and the *TM3, Sb* balancer were also collected in order to set up the next generation of crosses.

Next, 4-6 F1 males from each line were placed into two duplicate vials per line, for a total of 2-3 males per vial. These flies were backcrossed to 5 females per vial from each respective wild line. Parent flies were transferred to new vials after 2 days. F2 offspring were inspected using fluorescence microscopy under a Leica MZ16 fluorescent stereomicroscope. Offspring carrying the chromosome with the $INS^{C96Y}$ transgene were identified via the presence of GFP fluorescence in eyes. Four classes of offspring carrying the $INS^{C96Y}$ transgene were collected: males with the *TM3, Sb* balancer chromosome, males without *TM3, Sb*, females with *TM3, Sb*, and females without *TM3, Sb*. As with the F1 offspring, a goal was set of 5 flies per class per line, though in some lines it was impossible to collect 5 from each class due to decreased viability of flies with the *TM3, Sb* balancer. Collected flies were placed immediately into a freezer at -80°C.

*Thorax Measurements and Slide Preparation*

Flies were removed from the -80°C freezer by line and their thorax lengths were measured using a Nikon SMZ-2B microscope equipped with a mechanical stage and built-in micrometer (Fig. 3). They were then mounted on slides in order to be photographed. A thin layer of silicone vacuum grease was applied to glass slides. Whole flies were placed in the vacuum grease and their heads were oriented using a needle and syringe such that the entire surface area of one eye was visible facing directly upward. Two microcapillary tubes were placed around the flies to create a bridge, and halocarbon 700 oil was applied over the flies to reduce glare. A cover slip was then placed over the halocarbon oil and the two microcapillary tubes.

*Eye Photography and Image Analysis*

Eyes from all F1 flies and F2 flies from lines 174 through 195 were photographed using a Zeiss AxioCam HRc mounted on a Leica MZ16 fluorescent stereomicroscope. Images were captured and saved using the AxioVision 3.1.2.1 software. Eyes were photographed at 115X magnification. Eyes from F2 flies from lines 196 through 745 were photographed using a Leica DFC420 camera mounted on a Leica M205 FA automated fluorescence stereomicroscope at the same magnification. The Leica Application Suite software was used to merge z-stacks taken on the Leica M205 FA microscope and to analyze all images taken with both microscopes. An automatic analysis program was used to select the eye area and determine the area, major and minor axes, and aspect ratio for each eye photographed (Fig. 4). Eyes were scored by visual inspection for the presence or absence of lesions. A semi-quantitative scale was developed and used to score eyes for degree of ommatidial structure present, ranging from a 1 for nearly wildtype eyes to a 5 for eyes in which no ommatidia were visible. Eyes in which no eye tissue was visible were given a score of 6; however, these data points were excluded from analysis, as a lack of eye tissue was considered a lack of information on the extent of ommatidial degeneration.

*Statistical Analysis*

JMP statistical software version 8.0 was used to perform a statistical analysis of the data. A fully-nested, Model I ANOVA was performed on F1 data and F2 data separately for both eye area scaled by thorax length and aspect ratio (major axis length divided by minor axis length), with presence of *TM3, Sb* nested within line number and sex nested within presence of *TM3, Sb*. The ANOVA was also performed nesting presence of *TM3, Sb* within sex and sex within line number, but no significant difference was found between the two nesting schemes. A Student's *t*-test was performed on each analysis at the level of each factor (line number, presence of *TM3, Sb* within line number, and sex within presence of *TM3, Sb* within line number).

In order to analyze the presence or absence of lesions, which was treated as a binary, nominal variable, a contingency test was performed, comparing the prevalence of lesions within each line. To take into account the effect of sex and the presence or absence of *TM3, Sb*, a logistic regression was also performed on both the F1 and F2 data, using maximum likelihood to fit a linear model to the logistic response function. Line number, sex, and presence or absence of *TM3, Sb* were considered as covariates, but they were not nested. A modified logistic regression was performed on the F1 and F2 data for degree of ommatidial degeneration, which was treated as an ordinal variable with 5 response levels. In this case, the JMP software modeled the cumulative probability of being at or below each response level. A linear model with the same slope but a different intercept was fit to each of $r$-1 cumulative logistic response comparisons, with $r$ different response levels. Again, line number, sex, and presence or absence of *TM3, Sb* were considered as covariates but were not nested.

To compare the effect of homozygosity versus heterozygosity for a wild third chromosome, a nested Model I ANOVA was performed on scaled eye area and aspect ratio data from all female flies without the *TM3, Sb* balancer. Number of copies of the wild third chromosome was nested within line number. The same ANOVA was performed on data from all male flies without the *TM3, Sb* balancer, separately from the females. A Student's *t*-test was performed on the results from each of these analyses at the level of each factor (line number and number of copies of the wild third chromosome or generation within line number).

Microsoft Excel version 12.2.4 was used to generate scatter plots of distributions of quantitative aspects of eye phenotype.

**Results**

*Variation in eye size*

Background genetic variation was visible as a broad range of variation in eye size, with eyes ranging from nearly wild-type to highly reduced and slit-like (Fig. 5, Fig. 6). A bivariate fit of eye area versus thorax length did not show a significant correlation for F1 flies ($r^2 = 0.0051$). For F2 data, there was a weak correlation ($r^2 = 0.36$) (Fig. 7). In addition, a linear regression of thorax length against line showed no significant correlation for either F1 flies ($r^2=0.015$) or F2 flies ($r^2=0.000903$), demonstrating that line did not have a significant effect on thorax length. As thorax length has been used as a scaling, overall indicator of body size, the lack of a tight correlation between eye area and thorax length indicates that the variation we observed in eye size was not due to variation in body size overall. Nonetheless, we standardized eye size against thorax length in order to remove any possible confounding effect of body size. A Model I nested ANOVA was performed with eye size of F1 flies as the response variable and three nominal explanatory variables, line number, presence or absence of *TM3, Sb* nested within line number, and sex nested within presence or absence of *TM3, Sb*. The results showed significant variation between sex within presence or absence of *TM3, Sb* within line (F=47.6986, P<0.0001), significant variation between presence or absence of *TM3, Sb* within line (F=19.6633, P<0.0001), and significant variation among lines (F=21.2890, P<0.0001) (Table 1). The explanatory variables of the nested ANOVA model accounted for 87% of the observed variation between individual eyes, showing that the model fit the data well ($r^2 = 0.87$). A Student's *t*-test showed that the eye sizes fell into statistically distinguishable classes that nonetheless formed a continuum of phenotypes (Table 2).

The same nested ANOVA performed on the eye sizes of the F2 flies showed significant variation between sex within presence or absence of *TM3, Sb* within line (F=26.5208, P<0.0001), significant variation between presence or absence of *TM3, Sb* within line (F=16.0366, P<0.0001), and significant variation among lines (F=38.7096, P<0.0001) (Table 1). The ANOVA model explained the observed variation in F2 eye sizes similarly well as for the F1 data ($r^2 = 0.86$). A Student's *t*-test was also performed on the F2 data at each level of grouping. As with the F1 eye sizes, the F2 eye sizes fell into statistically distinguishable classes that formed a continuum of phenotypes overall (Table 3). In both F1 and F2 flies, line 199 was the line with the smallest eye size when the data were analyzed only by line. When the data were further broken down into presence or absence of *TM3, Sb* within line and sex within presence or absence of *TM3, Sb*, different classes of data appeared as having the smallest eye size. Among F1 flies, the class with the smallest eyes was male flies with *TM3, Sb* from line 176, whereas among F2 flies, the class with the smallest eyes was male flies with *TM3, Sb* from line 189. However, the same lines regularly appeared among classes with the smallest and largest eye sizes both in F1 and F2 flies.

*Variation in aspect ratio*

The aspect ratio of the eyes (length of major axis divided by length of minor axis) was used as a measure of eye shape. We observed by visual inspection that, in some lines, smaller eyes were narrow and elongated, resulting in a larger aspect ratio, whereas in other lines, smaller eyes retained roughly the same aspect ratio as larger eyes (Fig. 8). A Model I nested ANOVA was performed with the aspect ratio of F1 eyes as the response variable and three nominal explanatory variables, line number, presence or absence of *TM3, Sb*, and sex nested as before. The results showed significant variation between sex within presence or absence of *TM3, Sb* within line (F=14.1069, P<0.0001), significant variation between presence or absence of *TM3, Sb* within line (F=8.5930, P<0.0001), and significant variation among lines (F=5.6845, P<0.0001) (Table 1). The ANOVA model explained slightly less of the observed variation than it did for eye area data ($r^2 = 0.70$), though it still accounted for most of the variance in the trait; a larger spread was observed for data points with a higher aspect ratio. A Student's *t*-test performed on the data grouped by line, by presence or absence of *TM3, Sb* within line, and by sex within presence or absence of *TM3, Sb* showed that, as with eye size, the aspect ratios fell into statistically distinguishable classes that formed a continuum of phenotypes overall (Table 4). However, as a higher proportion of the data fell into classes with a smaller aspect ratio (closer to wild-type) than into classes with a larger aspect ratio, the classes with smaller aspect ratios showed a greater statistical overlap than the classes with larger aspect ratios.

The aspect ratios of F2 eyes displayed a greater spread than those of F1 eyes,

particularly among eyes with a larger aspect ratio. There were several outliers among the data, corresponding to extremely reduced, slit-like eyes. Outliers more than three standard deviations above the mean were removed from analysis. An ANOVA with the same nested factors as for the F1 data was performed on the F2 data. The results showed significant variation between sex within presence or absence of *TM3, Sb* within line (F=7.7914, P<0.0001), significant variation between presence or absence of *TM3, Sb* within line (F=4.2230, P<0.0001), and significant variation among lines (F=9.1941, P<0.0001) (Table 1). The model did not explain as great a percentage of the variation in the F2 data as for the F1 data ($r^2 = 0.63$); however, a clear correlation was still visible between predicted and observed values. As with the F1 data, a Student's *t*-test performed on the data grouped by line, by presence or absence of *TM3, Sb* within line, and by sex within presence or absence of *TM3, Sb* showed that the aspect ratios fell into statistically distinguishable classes forming a continuum overall (Table 5). Among both the F1 and F2 data, line 193 appeared as the line with the largest mean aspect ratio (greatest departure from wild-type eyes). Unlike with eye size, this held true even when the data were further broken down into presence or absence of *TM3, Sb* nested within line and sex nested within presence or absence of *TM3, Sb*. There was not one line that consistently appeared as the line with the smallest mean aspect ratio; however, several lines, such as 187 and 182, consistently fell out among the 7 to 8 lines with the smallest aspect ratios, which showed some statistical overlap. Although there was not a perfect correlation between eye size and aspect ratio for either F1 or F2 flies, the lines with the largest eyes tended to have smaller aspect ratios, corresponding to a less severe overall phenotype, whereas the lines with the smallest eyes tended to have larger aspect ratios.

*Presence and pattern of lesions*

In most of the lines, black lesions appeared in the eyes of some or all flies; however, we observed that they appeared with greater frequency in some lines than in others and that their size and position differed between lines (Fig. 9). As a preliminary analysis of this trait, every eye was scored for the presence or absence of lesions as a binary trait. A contingency test performed on F1 data comparing all lines showed that the prevalence of lesions was not independent of line number (P<0.0001). There was a strong effect of sex (for a 2-sided *t*-test, P<0.0001), with males being much more likely to show lesions than females, and a lesser but still significant effect of the presence or absence of *TM3, Sb* (for a 2-sided *t*-test, P=0.0003), with flies carrying the balancer being somewhat more likely to show lesions than flies not carrying it (Fig. 10). A logistic regression showed a highly significant effect of line ($\chi^2$=142.49, P<0.0001), presence or absence of *TM3, Sb,* ($\chi^2$=48.81, P<0.0001) and sex ($\chi^2$=218.22, P<0.0001) on the presence or absence of lesions (Table 6). A test of the whole model, taking into account all three covariates, was also highly significant ($\chi^2$=314.11, P<0.0001).

As with the F1 data, a contingency test performed on the F2 data also showed that the prevalence of lesions was not independent of line number ($\chi^2$=174.683, P<0.0001). There was also a highly significant effect of sex (for a 2-sided *t*-test, P<0.0001), with males being more likely to show lesions than females, and a highly significant effect of the presence or absence of *TM3, Sb* (for a 2-sided *t*-test, P<0.0001), with flies carrying the balancer chromosome being more likely to show lesions than flies not carrying it (Fig. 11). A logistic regression on the F2 data also showed a highly significant effect of line ($\chi^2$=219.80, P<0.0001), presence or absence of *TM3, Sb* ($\chi^2$=66.168, P<0.0001), and sex ($\chi^2$=48.894, P<0.0001) on the presence or absence of lesions (Table 6). Again, a test of the whole model was highly significant ($\chi^2$=290.111, P<0.0001).

To test whether the same genetic backgrounds that tended to cause a severe reduction in eye size also tended to cause lesions, the line numbers ordered by mean scaled eye area were plotted against the line numbers ordered by percentage of flies with lesions. A linear regression was run on the plots for both F1 and F2 flies, and neither showed any significant correlation (for F1, $r^2$=0.00787; for F2, $r^2$=0.01071).

*Variation in ommatidial degeneration*

We observed through visual inspection that there was a wide range of variation in the degree of ommatidial structure present among eyes. While the degree of ommatidial degeneration generally appeared to correlate with the severity of other measures of phenotype (eye size and aspect ratio), there was some variation in degeneration between eyes that were similarly distorted or reduced in size. In order to analyze this variation, a semi-quantitative scale of ommatidial degeneration was developed and all eyes were scored. A modified logistic regression was performed on both the F1 and F2 data. For the F1 data, a highly significant effect was observed for line ($\chi^2$=444.805, P<0.0001), presence or absence of *TM3, Sb* ($\chi^2$=364.241, P<0.0001), and sex ($\chi^2$=696.206, P<0.0001). For the F2 data, as well, a highly significant effect was observed for line ($\chi^2$=554.48, P<0.0001), presence or absence of *TM3, Sb* ($\chi^2$=198.187, P<0.0001), and sex ($\chi^2$=479.545, P<0.0001) (Table 6). When the whole models were tested, a highly significant effect was observed in both generations (for F1, $\chi^2$=886.561, P<0.0001; for F2, $\chi^2$=843.397, P<0.0001). A lack of fit test also indicated that, for both generations, the logistic regression model accounted for essentially all of the observed variation. For the F1 data, sex had the highest chi-squared value, indicating that it was the factor with the greatest effect on the degree of ommatidial degeneration; however, for the F2 data, line had a slightly higher chi-squared value than sex, indicating that it had a greater effect on the degree of ommatidial degeneration.

*Comparison of third-chromosome heterozygotes and homozygotes*

We were curious as to whether we could observe a significant difference in phenotype due to dominant versus recessive effects of the wild third chromosomes. In order to detect such an effect, we first compared eye size of F1 females without *TM3, Sb*, which had one wild third chromosome and one third chromosome from the *w1118* laboratory line, and of F2 females without *TM3, Sb,* which had two copies of a wild third chromosome (Fig. 12). A Model I nested ANOVA with two nominal explanatory variables, line number and number of wild third chromosomes nested within line number, revealed significant variation between heterozygotes and homozygotes within line (F=5.8137, P<0.0001) and between lines (F=19.8602, P<0.0001) (Table 7). The ANOVA model explained 73% of the observed variation between individual eyes ($r^2$ = 0.73), and a Student's *t*-test showed that the eye sizes fell into statistically distinguishable classes when organized by line and by number of wild third chromosome within line that, again, formed a continuum of phenotype overall (Table 8). The same nested ANOVA was performed for aspect ratio. The results also showed significant variation between heterozygotes and homozygotes within line (F=2.3897, P<0.0001) and between lines (F=4.7114, P<0.0001) (Table 7); however, the model was not able to account for as much of the observed variation, indicating a higher presence of uncontrolled variation ($r^2 = 0.43$). A Student's *t*-test showed a higher degree of phenotypic overlap between classes, although a continuum was still visible (Table 9). A linear regression was performed on average eye area for each line for all F1 flies versus all F2 flies, revealing a correlation with an $r^2$ value of 0.4895. When line 202, which appeared to be an outlier and only contained three data points for the F1 flies, was removed, the correlation was stronger ($r^2 = 0.744$).

Phenotypic differences between male heterozygotes and homozygotes for wild third chromosomes were also compared (Fig. 13). A Model I nested ANOVA with two nominal explanatory variables, line number and number of wild third chromosomes nested within line, was performed to compare eye sizes between F1 males without *TM3, Sb*, which had one copy of a wild third chromosome, and F2 males without *TM3, Sb*, which had two copies of a wild third chromosome. The results showed significant variation between homozygotes and heterozygotes within line (F=10.6886, P<0.0001) and between lines (F=31.1017, P<0.0001) (Table 7). The ANOVA model explained 85% of the observed variation ($r^2 = 0.85$). Also, when a Student's *t*-test was performed for each level of organization of the data, the eye sizes fell into statistically distinguishable classes by line or by line and number of wild third chromosomes (Table 10). For this test, the classes of data formed a particularly smooth and regular continuum of phenotypes. The same nested ANOVA was performed for aspect ratio. As with the female data, the ANOVA showed significant variation between heterozygotes and homozygotes within line (F=5.6374, P<0.0001) and between lines (F=13.5147, P<0.0001) (Table 7).

However, the model did not explain as high a percentage of the observed variation as it did for eye area ($r^2 = 0.71$). A Student's *t*-test showed a greater phenotypic overlap among lines with smaller aspect ratios, while lines with larger aspect ratios were more statistically distinguishable from each other (Table 11). A linear regression was performed on average aspect ratio for each line for all F1 flies versus all F2 flies, revealing a correlation with an $r^2$ value of 0.509.

In order to compare qualitative aspects of eye phenotype, a modified logistic regression was performed on the ommatidial structure scores using line and number of wild third chromosomes as covariates for both females and males without *TM3, Sb*. The regression for the female data showed a significant effect of both line ($\chi^2_{36}=347.164$, P<0.0001) and number of wild third chromosomes ($\chi^2_{1}=54.288$, P<0.0001) on the degree of ommatidial structure present. However, the regression for the male data showed a significant effect of line ($\chi^2_{36}=342.656$, P<0.0001), but not a significant effect of number of wild third chromosomes ($\chi^2_{1}=0.0439$, P=0.8341), on the degree of ommatidial degeneration. A linear regression was performed on the average ommatidial structure score for each line for all F1 flies versus all F2 flies, revealing a correlation with an $r^2$ value of 0.5024. In addition, a linear regression was performed on the proportion of eyes with lesions for each line for all F1 flies versus all F2 flies. In this case, the correlation was considerably weaker than for all other characters compared ($r^2 = 0.1528$).

**Discussion**

The broad variation in eye phenotypes we observed in both F1 and F2 flies demonstrates that the background genetic variation present in isogenic lines derived from individuals randomly sampled from a single wild population has a significant effect on the extent of damage caused by expression of $INS^{C96Y}$. For each of the quantitative characters we surveyed, we saw a smooth continuum of phenotypic states, as would be expected for a trait in which many genes were interacting to produce a phenotype. At the same time, we were able to divide the phenotypes by line into several statistically distinguishable groups, indicating that the lines with the most severe phenotypes showed a statistically significant difference from the lines with the least severe phenotypes. Interestingly, while similar groups of lines displayed the most severe and least severe phenotypes for most characters scored, indicating some degree of pleiotropy in the generation of the phenotype, the exact same lines did not fall out at the top or bottom of the continuum of severity for each character. In particular, lines that displayed a high frequency of necrotic lesions and lines that tended to have highly degenerate ommatidia were not necessarily the same lines that displayed the greatest reduction in eye size or the highest degree of eccentricity in eye shape. Such a lack of a strict correlation indicates that more than one developmental pathway or suite of genes may be involved in generating the disease phenotype. Independently segregating polymorphisms in a population that act within different pathways may thus have a combinatorial effect on the final phenotype.

    The primary source of variance among both F1 flies and F2 flies was line number, as would be expected if differences in genetic background did in fact generate variance in severity of the disease phenotype. However, we also observed a statistically significant effect of both the presence or absence of the *TM3, Sb* balancer on the third chromosome and of sex. The *TM3, Sb* chromosome caused a broader spread of phenotypes overall,

causing both the least severe and most severe scores to take on more extreme values, though this spread was skewed toward more severe phenotypes. While this result was not expected, we hypothesize that *TM3, Sb* could exaggerate the $INS^{C96Y}$ phenotype simply through leading to the transcription and translation of a greater load of unstable, misfolded proteins. *TM3* contains various inversions, which are likely to disrupt coding sequences of proteins (Micol and García-Bellido, 1988). In addition, because balancer chromosomes are somewhat shielded from selective pressure, as they almost never occur in a homozygous state, they have the potential to accumulate deleterious recessive mutations relatively easily. Such mutations could translate into a higher level of unstable proteins than normal, which would place stress on ER chaperones and other pathways involved in protein folding and trafficking. Introducing the $INS^{C96Y}$ transgene into such a background could cause the cells' capability for dealing with unfolded proteins to be more easily exceeded, possibly leading to a greater incidence of apoptosis and a more severe phenotype. These results indicate that, in humans, *de novo* deleterious mutations like those that may accumulate on balancer chromosomes, although they are expected to be at a low frequency in the population, may have an important effect on the disease phenotype in some cases.

   Overall, male flies displayed more severe phenotypes than female flies, although the magnitude of this sex-specific effect varied by line. Sex-specific differences are not typically observed in *Drosophila* eye development, making this effect somewhat unexpected. However, studies of many different quantitative traits in *Drosophila melanogaster* have revealed a high degree of QTLs with sex-specific effects (Mackay, 2010). Because we have examined various dimensions of the eye disruption phenotype and because eye development is itself a very complex process, involving multiple sets of genes at different times throughout development, it is likely that we have uncovered alleles of certain genes active either in the developmental process or in protein folding and trafficking that differentially affect male and female flies. It is unclear whether this sex-specific effect is relevant to the biology of neonatal diabetes in other organisms; however, in a screening for *INS* mutations in human neonatal diabetes patients, it was found that, among carriers of *INS* mutations, birth weight was more reduced in males than in females (Edghill et al., 2008). Additionally, in the *Akita* mouse model, in which the C96Y mutation has occurred independently in the mouse *Ins2* gene, male mice display a more severe phenotype than female mice (Støy, 2007). Regardless of whether these effects are due to similar mechanisms, the sex-specificity of the pathogenesis of neonatal diabetes due to *INS* mutations will be interesting to study in greater detail in the future.

The appearance of black lesions and the variation in their prevalence and size was a particularly interesting and unexpected aspect of the *INS$_{C96Y}$* phenotype. Similar lesions have been described in other studies involving eye degeneration in *Drosophila* (Wang, et al., 2006; Bilen and Bonini, 2007; Lessing and Bonini, 2008). However, the exact cause of such lesions has not been characterized. If expression of *INS$^{C96Y}$* caused cell death by apoptosis at an early stage of eye development, the dead cells should not be visible as black lesions, as they would have been phagocytized by other cells. Such early apoptosis may be the cause of eyes with a reduced size. It is possible that the lesions could be caused by induced apoptosis taking place after the majority of eye development has occurred, due to activation of the unfolded protein response. Additionally, studies have indicated that pathways other than the unfolded protein response, such as oxidative stress, can cause lesions due to apoptosis in Drosophila eyes (Wang, et al., 2006; Steiner, et al., 2009).

We observed two main types of lesions: small lesions, about the size of one ommatidium, that often occurred repeatedly in the same eye, and large lesions that took up a significant proportion of the eye area, sometimes extending across almost the entire visible field of eye tissue. Small lesions were often present at the anterior margin of the eye, though they could be spread throughout. Large lesions, when they did not cover the majority of the eye, often occurred around a spot in the dorsal half of the eye, slightly toward the anterior; however, this pattern did not always hold true. These two types of lesions each occurred with a higher prevalence in certain lines, and they did not both tend to occur in flies within the same line. Such a separation of phenotype by line indicates that different polymorphisms within the population, possibly in genes acting at different stages of eye development, are likely to cause the two different types of lesions. In addition, the lines displaying a higher prevalence of either type of lesion did not tend to correspond to the lines with the greatest eye size reduction, distortion of eye shape, or ommatidial degeneration. This lack of correspondence suggests that generation of lesions is an aspect of the $INS^{C96Y}$ phenotype under independent genetic influence from eye reduction and ommatidial degeneration, rather than being caused by a limited set of pleiotropic genes. The genetics underlying the production of and variance in the $INS^{C96Y}$ phenotype thus appears to be complex in nature, capable of varying in multiple dimensions.

We were interested in comparing severity of phenotype between F1 flies with one copy of a wild third chromosome and corresponding F2 flies with two copies of a wild chromosome. If dominant effects existed in the genetic variation we sampled, they would be visible in the heterozygous F1 flies, whereas recessive effects would only be visible in homozygous F2 flies. The fact that line number accounted for the primary source of phenotypic variance in individuals both heterozygous and homozygous for a wild third chromosome shows that we uncovered both dominant and recessive effects of genetic variation. In the analyses of variance for eye size and aspect ratio, number of wild third chromosome was also a significant source of variance. This result indicates that whether a fly is heterozygous or homozygous for a wild third chromosome has a significant effect on the production of those traits. In both males and females, it appeared that recessive effects, which could only be observed in flies homozygous for a wild third chromosome, caused a broadening of the range of phenotypic values. Rather than causing phenotypes to be more or less severe overall, having two copies of a wild third chromosome magnified the differences due to background genetic variation already visible in the phenotypes of individuals with only one copy. The fact that the correlations performed on F1 versus F2 mean phenotypic values for eye area, aspect ratio, and ommatidial structure score tended to have $r^2$ values around 0.5 indicates that, while there is likely some additive effect of allelic copy number on the phenotype, there is also likely an important contribution of recessive alleles. The correlation for proportions of F1 eyes with lesions versus proportions of F2 eyes with lesions was much lower; however, another measurement of the lesions, such as number of lesions per eye or percentage of eye area taken up by lesions, might yield more information.

    One drawback to our experimental design that we hope to improve upon in future

experiments was the use of the *UAS/GAL4* system. It is possible that some of the variation in phenotype that we observed was due to the effect of genetic variation on the efficiency of the *UAS/GAL4* system and the resulting transcription of $INS^{C96Y}$. Some lines may have genetically-based differences in the efficacy of the *GMR* promoter that we used to drive *GAL4* expression in the eyes. Additionally, differences in the very genes we expect to be relevant to the $INS^{C96Y}$ phenotype, those involved in protein folding and trafficking, may affect the ability of the GAL4 protein to activate the *UAS* enhancer and drive transcription of $INS^{C96Y}$. Nonetheless, based on our results, we expect variation in the efficiency of the *UAS/GAL4* system to have a minimal effect when compared to variation in genes that directly interact with the mutant insulin protein. The diversity of different phenotypes observed and particularly the apparent independence of some aspects of observed eye degeneration are difficult to explain simply based on different levels of $INS^{C96Y}$ transcription. Rather, they indicate that several different processes are at work in generating the $INS^{C96Y}$ phenotype and are significantly affected by background genetic variation. Still, because the *UAS/GAL4* system has the potential to introduce unwanted variance into our data, possibly obscuring the true effect of background genetic variation on the production of the disease phenotype, a line of flies with $INS^{C96Y}$ transcription driven directly by *GMR* would be ideal for future studies.

    As we have now established that background genetic variation has a significant effect on the severity of the eye degeneration phenotype produced by the expression of $INS^{C96Y}$ in fly eyes, we plan to explore several directions in future research. In order to better understand the kinds of genetic variation contributing to the observed phenotypic variation and to begin to identify relevant genes or genetic pathways, we plan to perform

a QTL analysis. It will be interesting to determine whether different loci contribute significantly to variation in the different aspects of the phenotype we measured, particularly those that did not appear tightly correlated. Because the genomes of the 36 wild lines we used have been sequenced, we can also perform a genome-wide association study. However, in order to increase the power of such a study, the number of lines involved would need to be significantly expanded. Because a QTL analysis may detect rare polymorphisms that happen to be present in the individuals chosen for the study, whereas an association study identifies more common polymorphisms that are segregating in the population, these two types of study could provide useful and complementary information.

Another possible future experiment we hope to undertake involves using artificial selection on lines with more or less severe phenotypes to try to drive them farther apart phenotypically. In addition to creating more extreme phenotypes, we expect that such selection would cause a change in allelic frequencies, favoring those alleles that contribute toward creating either a more or less severely disrupted eye. By analyzing how the allelic frequencies of the population under selection change and which genes show the greatest response to selection, we can better understand the complex genetic architecture underlying the production of the degenerate eye phenotype in response to the expression of $INS^{C96Y}$.

This study provides an important first step for the future analysis of the genetics of neonatal diabetes. We can conclude from our results that the phenotypes caused by insulin gene mutations are highly sensitive to background genetic variation. We have shown statistically significant variation of eye area, eye shape, presence of lesions, and degree of ommatidial degeneration in eyes of *Drosophila* expressing $INS^{C96Y}$ in different genetic backgrounds. Studies of human neonatal diabetes patients carrying insulin gene mutations have reported significant variance in penetrance of the disease both within affected families and between families carrying the same mutation (Edghill, et al., 2008). The evolutionary conservation of many fundamental elements of the protein homeostasis pathways and the unfolded protein response between *Drosophila* and mammals indicates that a more in-depth analysis of our $INS^{C96Y}$ fly lines has the potential to shed light on the pathogenesis of permanent neonatal diabetes due to *INS* mutations in humans. Based on the complexity of the relevant genetic pathways, we expect that future genetic studies of neonatal diabetes in multiple organisms will reveal that a large number of genes, acting both additively and epistatically, can influence the severity and penetrance of the disease phenotype. If this proves true, further studies in *Drosophila* will offer a level of power and experimental versatility difficult to achieve in mammalian models.

The sequencing of the human genome and the advent of genomics-based medicine offer the promise of much more sophisticated methods for treating disease. However, the realization that the majority of genetic diseases cannot be simply explained by mutations in a single gene has complicated the panorama. In order to develop truly individualized, effective screens and treatments for complex genetic diseases, researchers and doctors may have to take into account not one or a few genes, but an entire genome's worth of polymorphisms and background genetic variation. Such treatment may, in fact, require a shift in the paradigm of understanding genetic diseases. As with any other trait displaying a significant range of quantitative phenotypic variation, for most genetic diseases, it is likely that any individual phenotype is caused by interactions between many genes involved in diverse pathways. For early-onset diseases such as neonatal diabetes, the spatial and temporal complexity of the developmental program itself is also fundamentally relevant to the production of the disease phenotype. We anticipate that future studies building on our results will both uncover genes and pathways with the greatest contribution to variance in the severity and penetrance of permanent neonatal diabetes and begin to elaborate a framework for the treatment of complex genetic diseases within the context of individual genetic background.

**Acknowledgements**

Many thanks to the entire Kreitman lab for their constant support. In particular, thanks to Emily Hudson for keeping the wild lines and measuring so many thoraxes. Thanks also to Levi Barse for helping with thorax measurements and spending so many hours in front of a computer analyzing images and formatting Excel spreadsheets. Thanks to Soo-Young Park in the Bell lab for helping create the transgenic flies. Thanks to Dr. Anna Pluzhnikov for statistical consultations. Special thanks to Misha, Marty, and Cecelia, for their confidence in me, for providing me with many thought-provoking conversations, and for teaching me so much about how to be a scientist.

**Figure Legends**

**Figure 1.** Diagram of the preproinsulin molecule, indicating the wild-type folded structure, the signal peptide (in green), the A chain (in blue), the B chain (in red), the C-peptide (in orange), the position of disulfide bridges, and the position of the C96Y mutation (in black). After Støy, et al., 2007.

**Figure 2.** Crossing scheme. Chromosomes labeled *w1118* correspond to the *w1118* laboratory line; chromosomes labeled *i* correspond to the *i*th wild line.

**Figure 3.** Illustration of thorax measurements. The black arrow indicates the axis that was used to measure thorax length for all flies.

**Figure 4.** Illustration of the process used to generate eye area and aspect ratio. The eye area identified by the analysis program was converted to a solid shape (green pixels). The program counted total green pixels to measure eye area and measured the major and minor axes of the green shape to calculate aspect ratio. (A) Photograph of an eye close to wild-type shape. (B) Eye area selected by analysis program for eye in part A. (C) A more severely reduced eye. (D) Eye area selected by analysis program for eye in part C. Note that the analysis program was capable of identifying small irregularities in the outline of the eye.

**Figure 5.** Range of eye sizes. (A, B, C, D) Photographs representative of range of different eye sizes observed (A, least severe; D, most severe).

**Figure 6.** Distribution of all scaled eye areas. (A) Scatter plot showing range of all measured F1 eye areas, scaled against thorax length. Eyes are ordered on the x-axis ascending by area. (B) Scatter plot showing range of all measured F2 eye areas, scaled

against thorax length. Eyes are ordered on the x-axis ascending by area.

**Figure 7.** Bivariate fit of unscaled eye area versus thorax length. (A) There is no significant correlation between unscaled eye area and thorax length in F1 flies ($r_2$=0.0051). (B) There is a very low correlation between unscaled eye area and thorax length in F2 flies ($r_2$=0.36).

**Figure 8.** Range of eye aspect ratios. (A) Scatter plot showing the range of all measured F1 aspect ratios. Eyes are ordered on the x-axis by ascending aspect ratio. (B) Scatter plot showing the range of all measured F2 aspect ratios. Eyes are ordered on the x-axis by ascending aspect ratio.

**Figure 9.** Types of lesions seen in fly eyes. (A) Small lesions, spread throughout eye. (B) Larger lesion, located in anterior, dorsal quadrant of eye. A few small lesions are also visible along the posterior, ventral edge. (C) Largest lesion, covering nearly entire eye surface.

**Figure 10.** Mosaic plots of distribution of lesions in F1 flies. For all plots, red bars represent proportion of flies with lesions, and blue bars represent proportion of flies without lesions. The right-most bars in each plot show proportion of total flies with (red) or without (blue) lesions. (A) Proportion of flies with or without lesions in each line. (B) Proportion of flies with or without lesions by sex. (C) Proportion of flies with or without lesions by presence or absence of *TM3, Sb*.

**Figure 11.** Mosaic plots of distribution of lesions in F2 flies. For all plots, red bars represent proportion of flies with lesions, and blue bars represent proportion of flies without lesions. The right-most bars in each plot show proportion of total flies with (red) or without (blue) lesions. (A) Proportion of flies with or without lesions in each line. (B)

Proportion of flies with or without lesions by sex. (C) Proportion of flies with or without lesions by presence or absence of *TM3, Sb.*

**Figure 12.** Distribution of quantitative aspects of eye phenotype for females heterozygous for a wild third chromosome versus females homozygous for a wild third chromosome. (A) Scatter plot of distribution of scaled eye areas. Eyes are ordered on the x-axis ascending by area. (B) Scatter plot of distribution of aspect ratios. Eyes are ordered on the x-axis ascending by ratio.

**Figure 13.** Distribution of quantitative aspects of eye phenotype for males heterozygous for a wild third chromosome versus males homozygous for a wild third chromosome. (A) Scatter plot of distribution of scaled eye areas. Eyes are ordered on the x-axis ascending by area. (B) Scatter plot of distribution of aspect ratios. Eyes are ordered on the x-axis ascending by ratio.

**Tables**

**Table 1.** Quantitative eye measurement traits (size and shape) in F1 and F2 flies analyzed with Model I, fully-nested ANOVA.

|  | Source | df[a] | $F$[b] | $P$[c] |
|---|---|---|---|---|
| F1 Scaled eye area | Line | 35 | 21.289 | <0.0001 |
|  | *TM3, Sb* (Line) | 36 | 19.6633 | <0.0001 |
|  | Sex (*TM3, Sb,* Line) | 52 | 47.6986 | <0.0001 |
| F2 Scaled eye area | Line | 33 | 38.7096 | <0.0001 |
|  | *TM3, Sb* (Line) | 34 | 16.0366 | <0.0001 |
|  | Sex (*TM3, Sb,* Line) | 63 | 26.5208 | <0.0001 |
| F1 Aspect ratio | Line | 35 | 5.6845 | <0.0001 |
|  | *TM3, Sb* (Line) | 36 | 8.593 | <0.0001 |
|  | Sex (*TM3, Sb,* Line) | 52 | 114.1069 | <0.0001 |
| F2 Aspect ratio | Line | 33 | 9.1941 | <0.0001 |
|  | *TM3, Sb* (Line) | 34 | 4.223 | <0.0001 |
|  | Sex (*TM3, Sb,* Line) | 63 | 7.7914 | <0.0001 |

[a] Degrees of freedom

[b] F statistic

[c] Probability F>F$_{obs.}$

**Table 2.** Results from a Student's *t*-test on F1 scaled eye areas analyzed by line. Levels not connected by the same letter are significantly different.

| Level | | Least Sq Mean |
|---|---|---|
| 202 | A | 114382.97 |
| 184 | B | 91698.84 |
| 190 | B C | 87496.97 |
| 209 | C D | 79870.38 |
| 182 | D | 79413.05 |
| 194 | D E | 77367.14 |
| 745 | D E F | 75095.44 |
| 179 | D E F | 74391.71 |
| 177 | D E F G H | 72420.30 |
| 445 | D E F G | 72337.54 |
| 187 | E F G H I | 70068.82 |
| 205 | E F G H I | 69462.29 |
| 185 | F G H I | 69327.39 |
| 183 | F G H I J | 67979.39 |
| 197 | F G H I J | 66189.39 |
| 186 | G H I J | 65781.45 |
| 175 | G H I J | 64906.26 |
| 200 | H I J | 64737.84 |
| 189 | H I J K | 63242.67 |
| 206 | H I J K L | 63141.29 |
| 201 | I J K | 63040.65 |
| 196 | I J K | 62570.18 |
| 207 | J K L M | 58240.12 |
| 176 | K L M | 55728.46 |
| 198 | K L M | 55231.26 |
| 188 | L M | 54811.49 |
| 195 | M | 53293.56 |
| 208 | M N | 51335.42 |
| 203 | M N | 50510.01 |
| 191 | M N | 50434.53 |
| 181 | N O | 42054.79 |
| 174 | O P | 39501.80 |
| 744 | O P | 39498.84 |
| 193 | O P | 38210.84 |
| 204 | O P | 36888.99 |
| 199 | P | 29230.98 |

**Table 3.** Results from a Student's *t*-test on F2 scaled eye areas analyzed by line. Levels not connected by the same letter are significantly different.

| Level | | | | | | | | | | | | | | | | | | | Least Sq Mean |
|---|---|---|---|---|---|---|---|---|---|---|---|---|---|---|---|---|---|---|---|
| 187 | A | | | | | | | | | | | | | | | | | | 98029.515 |
| 177 | A | | | | | | | | | | | | | | | | | | 92683.963 |
| 179 | | B | | | | | | | | | | | | | | | | | 84603.251 |
| 182 | | B | C | | | | | | | | | | | | | | | | 84104.451 |
| 190 | | B | C | D | | | | | | | | | | | | | | | 80221.756 |
| 445 | | | C | D | E | | | | | | | | | | | | | | 76640.588 |
| 186 | | | | D | E | | | | | | | | | | | | | | 75926.103 |
| 200 | | | | D | E | F | | | | | | | | | | | | | 74032.734 |
| 194 | | | | | E | F | G | | | | | | | | | | | | 71276.512 |
| 175 | | | | | E | F | G | H | | | | | | | | | | | 70261.750 |
| 209 | | | | | E | F | G | H | I | | | | | | | | | | 68414.042 |
| 180 | | | | | | F | G | H | I | | | | | | | | | | 67899.273 |
| 201 | | | | | | F | G | H | I | | | | | | | | | | 67041.662 |
| 205 | | | | | | F | G | H | I | | | | | | | | | | 66889.200 |
| 183 | | | | | | | G | H | I | J | | | | | | | | | 64711.756 |
| 745 | | | | | | | G | H | I | J | | | | | | | | | 63454.662 |
| 197 | | | | | | | | H | I | J | | | | | | | | | 63111.699 |
| 176 | | | | | | | | H | I | J | | | | | | | | | 63092.539 |
| 198 | | | | | | | | H | I | J | | | | | | | | | 63080.210 |
| 206 | | | | | | | | | I | J | | | | | | | | | 61702.710 |
| 185 | | | | | | | | | | J | K | | | | | | | | 57357.560 |
| 196 | | | | | | | | | | | K | L | | | | | | | 53058.679 |
| 195 | | | | | | | | | | | K | L | | | | | | | 52826.211 |
| 188 | | | | | | | | | | | K | L | M | | | | | | 49944.407 |
| 202 | | | | | | | | | | | | L | M | | | | | | 49330.938 |
| 174 | | | | | | | | | | | | L | M | N | | | | | 43844.719 |
| 192 | | | | | | | | | | | | | M | N | | | | | 42997.326 |
| 203 | | | | | | | | | | | | | M | N | O | | | | 41271.390 |
| 744 | | | | | | | | | | | | | | N | O | P | | | 36592.743 |
| 189 | | | | | | | | | | | | | | N | O | P | Q | | 36503.924 |
| 191 | | | | | | | | | | | | | | | O | P | Q | | 31363.998 |
| 181 | | | | | | | | | | | | | | | | P | Q | | 30883.288 |
| 193 | | | | | | | | | | | | | | | | | Q | | 27124.709 |
| 199 | | | | | | | | | | | | | | | | | | R | 15191.960 |

**Table 4.** Results from a Student's *t*-test on F1 aspect ratios analyzed by line. Levels not connected by the same letter are significantly different.

| Level | | Least Sq Mean |
|---|---|---|
| 193 | A | 1.8844001 |
| 199 | A B | 1.7762421 |
| 201 | B | 1.7439559 |
| 204 | B C | 1.7391784 |
| 744 | B C | 1.7367513 |
| 174 | B C D | 1.7038869 |
| 177 | B C D E | 1.7029444 |
| 176 | B C D | 1.7016566 |
| 198 | B C D E | 1.6937882 |
| 181 | B C D E | 1.6897435 |
| 203 | B C D E | 1.6884935 |
| 445 | C D E F | 1.6588334 |
| 196 | C D E F G | 1.6545713 |
| 205 | C D E F G H | 1.6530147 |
| 206 | C D E F G H I | 1.6471872 |
| 200 | D E F G H I | 1.6367549 |
| 185 | D E F G H I J | 1.6246652 |
| 175 | D E F G H I J K | 1.6218538 |
| 191 | D E F G H I J K | 1.6182410 |
| 197 | E F G H I J K L | 1.6080688 |
| 183 | F G H I J K L | 1.6001265 |
| 194 | F G H I J K L M | 1.5975017 |
| 188 | G H I J K L M N | 1.5833192 |
| 189 | F G H I J K L M N | 1.5831632 |
| 208 | F G H I J K L M N | 1.5800056 |
| 186 | F G H I J K L M N | 1.5776314 |
| 195 | H I J K L M N | 1.5720717 |
| 187 | I J K L M N | 1.5601272 |
| 182 | J K L M N | 1.5520981 |
| 207 | I J K L M N | 1.5415834 |
| 190 | K L M N | 1.5388303 |
| 745 | K L M N | 1.5345776 |
| 184 | L M N | 1.5247682 |
| 179 | N | 1.5106316 |
| 209 | N | 1.5092603 |
| 202 | M N | 1.4431706 |

**Table 5.** Results from a Student's *t*-test on F2 aspect ratios analyzed by line. Levels not connected by the same letter are significantly different.

| Level | | Least Sq Mean |
|---|---|---|
| 193 | A | 1.9155591 |
| 191 | A B | 1.8688945 |
| 192 | A B | 1.8372267 |
| 199 | B C | 1.8089238 |
| 174 | B C D | 1.7446108 |
| 181 | C D E | 1.7158774 |
| 203 | D E F | 1.6924428 |
| 176 | D E F | 1.6835043 |
| 183 | D E F | 1.6798175 |
| 744 | D E F | 1.6781717 |
| 206 | D E F | 1.6746936 |
| 175 | D E F | 1.6651677 |
| 188 | D E F G | 1.6603013 |
| 209 | D E F G H | 1.6481442 |
| 201 | D E F G H | 1.6472596 |
| 445 | D E F G H I | 1.6456866 |
| 185 | D E F G H | 1.6438873 |
| 205 | E F G H I | 1.6342538 |
| 196 | E F G H I J | 1.6304111 |
| 189 | D E F G H I J K L M | 1.6182016 |
| 198 | F G H I J K | 1.6173639 |
| 194 | F G H I J K L | 1.6125756 |
| 197 | G H I J K L M | 1.5784960 |
| 202 | G H I J K L M N | 1.5703306 |
| 180 | H I J K L M N | 1.5668704 |
| 179 | I J K L M N | 1.5606298 |
| 190 | J K L M N | 1.5439588 |
| 195 | K L M N | 1.5307369 |
| 745 | K L M N | 1.5258276 |
| 177 | L M N | 1.5242598 |
| 200 | M N | 1.5212713 |
| 182 | M N | 1.5180935 |
| 186 | N O | 1.4795702 |
| 187 | O | 1.4113303 |

**Table 6.** Nominal and ordinal eye traits in F1 and F2 flies analyzed with a logistic regression.

|  | Source | df[a] | $\chi^2$ | $P$[b] |
|---|---|---|---|---|
| F1 Lesions | Line | 35 | 142.495 | <0.0001 |
|  | *TM3, Sb* | 1 | 48.8098 | <0.0001 |
|  | Sex | 1 | 218.222 | <0.0001 |
| F2 Lesions | Line | 33 | 219.8 | <0.0001 |
|  | *TM3, Sb* | 1 | 66.168 | <0.0001 |
|  | Sex | 1 | 48.894 | <0.0001 |
| F1 Structural score | Line | 35 | 444.805 | <0.0001 |
|  | *TM3, Sb* | 1 | 364.241 | <0.0001 |
|  | Sex | 1 | 696.206 | <0.0001 |
| F2 Structural score | Line | 33 | 554.48 | <0.0001 |
|  | *TM3, Sb* | 1 | 198.187 | <0.0001 |
|  | Sex | 1 | 479.545 | <0.0001 |

[a] Degrees of freedom

[b] Probability $\chi^2 > \chi^2_{obs}$.

**Table 7.** Quantitative eye measurement traits (size, shape) in female and male flies heterozygous and homozygous for a wild third chromosome analyzed with a Model I, fully-nested ANOVA.

|  | Source | df[a] | $F^b$ | $P^c$ |
|---|---|---|---|---|
| Female scaled eye area | Line | 37 | 19.8602 | <0.0001 |
|  | # Wild 3rd chromosomes (Line) | 31 | 5.8137 | <0.0001 |
| Female aspect ratio | Line | 37 | 4.7114 | <0.0001 |
|  | # Wild 3rd chromosomes (Line) | 31 | 2.3897 | <0.0001 |
| Male scaled eye area | Line | 36 | 31.1017 | <0.0001 |
|  | # Wild 3rd chromosomes (Line) | 27 | 10.6886 | <0.0001 |
| Male aspect ratio | Line | 36 | 13.5147 | <0.0001 |
|  | # Wild 3rd chromosomes (Line) | 27 | 5.6374 | <0.0001 |

[a] Degrees of freedom

[b] F statistic

[c] Probability F>F$_{obs.}$

**Table 8.** Results of a Student's *t*-test on scaled eye area from female flies both heterozygous and homozygous for a wild third chromosome. Levels not connected by the same letter are significantly different.

| Level | | | | | | | | | | | | | | | | | | Least Sq Mean |
|---|---|---|---|---|---|---|---|---|---|---|---|---|---|---|---|---|---|---|
| 177 | A | | | | | | | | | | | | | | | | | 119053.30 |
| 202 | A | B | C | | | | | | | | | | | | | | | 108212.21 |
| 179 | | B | | | | | | | | | | | | | | | | 107339.51 |
| 187 | | B | C | | | | | | | | | | | | | | | 104969.06 |
| 182 | | B | C | | | | | | | | | | | | | | | 104556.79 |
| 209 | | B | C | | | | | | | | | | | | | | | 103637.21 |
| 186 | | B | C | | | | | | | | | | | | | | | 102735.10 |
| 184 | | B | C | D | | | | | | | | | | | | | | 102666.11 |
| 200 | | B | C | D | | | | | | | | | | | | | | 101814.97 |
| 201 | | B | C | D | E | | | | | | | | | | | | | 99579.13 |
| 445 | | B | C | D | | | | | | | | | | | | | | 99481.02 |
| 194 | | B | C | D | E | | | | | | | | | | | | | 98922.61 |
| 190 | | B | C | D | E | | | | | | | | | | | | | 98585.44 |
| 180 | | B | C | D | E | F | | H | | | | | | | | | | 97335.02 |
| 205 | | | C | D | E | F | G | | | | | | | | | | | 95365.81 |
| 183 | | | C | D | E | F | G | H | | | | | | | | | | 95303.32 |
| 185 | | | | D | E | F | G | H | I | | | | | | | | | 92677.15 |
| 745 | | B | C | D | E | F | G | H | I | J | K | L | | | | | | 90649.44 |
| 197 | | | | | E | F | G | H | I | J | | | | | | | | 89698.14 |
| 196 | | | | | | F | G | H | I | J | K | | | | | | | 86773.29 |
| 195 | | | | | | F | G | H | I | J | K | | | | | | | 86109.89 |
| 198 | | | | | | | | H | I | J | K | L | | | | | | 85633.22 |
| 188 | | | | | | | | H | I | J | K | L | | | | | | 85479.18 |
| 189 | | | | | | | | H | I | J | K | L | | | | | | 85415.14 |
| 208 | | | | | | F | G | H | I | J | K | L | | | | | | 84141.28 |
| 176 | | | | | | | | | I | J | K | L | | | | | | 83706.75 |
| 175 | | | | | | | | | I | J | K | L | | | | | | 82950.63 |
| 192 | | | | | | | G | | I | J | K | L | M | N | | | | 80292.44 |
| 207 | | | | | | | | | | J | K | L | M | | | | | 78525.18 |
| 203 | | | | | | | | | | | K | L | | | | | | 77216.98 |
| 206 | | | | | | | | | | | | L | M | | | | | 75875.95 |
| 193 | | | | | | | | | | | | | M | N | O | | | 66985.86 |
| 181 | | | | | | | | | | | | | | N | O | | | 65757.10 |
| 744 | | | | | | | | | | | | | | N | O | | | 65726.45 |
| 174 | | | | | | | | | | | | | | N | O | | | 65453.95 |
| 204 | | | | | | | | | | | | | | | O | P | | 62272.27 |
| 191 | | | | | | | | | | | | | | | | P | | 55752.15 |
| 199 | | | | | | | | | | | | | | | | | Q | 36459.90 |

**Table 9.** Results of a Student's *t*-test on aspect ratio from female flies both heterozygous and homozygous for a wild third chromosome. Levels not connected by the same letter are significantly different.

| Level | | Least Sq Mean |
|---|---|---|
| 199 | A | 1.6826260 |
| 192 | A B | 1.6730130 |
| 191 | A B | 1.6675571 |
| 193 | A B C | 1.6089162 |
| 176 | A B C D | 1.5986031 |
| 204 | A B C D E | 1.5904769 |
| 175 | B C D E | 1.5840116 |
| 189 | C D E F | 1.5447984 |
| 206 | C D E F | 1.5442049 |
| 207 | C D E F G | 1.5345765 |
| 445 | C D E F G | 1.5262994 |
| 203 | C D E F G | 1.5258443 |
| 174 | C D E F G | 1.5255148 |
| 744 | C D E F G | 1.5236915 |
| 196 | D E F G | 1.5185976 |
| 198 | E F G H | 1.5112896 |
| 181 | E F G | 1.5102741 |
| 201 | E F G H I | 1.5091757 |
| 188 | E F G H I | 1.5071355 |
| 197 | E F G H I J | 1.4997588 |
| 183 | E F G H I J | 1.4960800 |
| 190 | F G H I J | 1.4913797 |
| 202 | C D E F G H I J K | 1.4884932 |
| 195 | F G H I J K | 1.4763741 |
| 194 | F G H I J K | 1.4759391 |
| 185 | F G H I J K | 1.4668857 |
| 179 | G H I J K | 1.4520384 |
| 180 | F G H I J K | 1.4518056 |
| 205 | G H I J K | 1.4508922 |
| 209 | G H I J K | 1.4440803 |
| 177 | G H I J K | 1.4408400 |
| 208 | F G H I J K | 1.4359420 |
| 186 | H I J K | 1.4289291 |
| 200 | I J K | 1.4234511 |
| 187 | J K | 1.4168533 |
| 184 | J K | 1.4061584 |
| 182 | K | 1.4023656 |

**Table 10.** Results of a Student's *t*-test on scaled eye area from male flies both heterozygous and homozygous for a wild third chromosome. Levels not connected by the same letter are significantly different.

| Level | | | | | | | | | | | | | | | | | | Least Sq Mean |
|---|---|---|---|---|---|---|---|---|---|---|---|---|---|---|---|---|---|---|
| 182 | A | | | | | | | | | | | | | | | | | 83355.201 |
| 179 | A | | | | | | | | | | | | | | | | | 81082.151 |
| 177 | A | B | | | | | | | | | | | | | | | | 76559.681 |
| 200 | | B | C | | | | | | | | | | | | | | | 70426.023 |
| 187 | | B | C | | | | | | | | | | | | | | | 70014.849 |
| 180 | | B | C | D | | | | | | | | | | | | | | 67567.723 |
| 445 | | | C | D | E | | | | | | | | | | | | | 63885.440 |
| 196 | | | C | D | E | | | | | | | | | | | | | 62043.320 |
| 745 | | | | D | E | | | | | | | | | | | | | 61065.199 |
| 185 | | | | D | E | F | | | | | | | | | | | | 59900.351 |
| 201 | | | | | E | F | G | | | | | | | | | | | 57420.791 |
| 194 | | | | | E | F | G | H | | | | | | | | | | 57321.875 |
| 183 | | | | | E | F | G | H | | | | | | | | | | 57130.942 |
| 190 | | | | | E | F | G | H | | | | | | | | | | 56476.349 |
| 195 | | | | | E | F | G | H | I | | | | | | | | | 55638.048 |
| 209 | | | | | E | F | G | H | I | J | | | | | | | | 54486.060 |
| 176 | | | | | | F | G | H | I | J | | | | | | | | 52669.672 |
| 175 | | | | | | | G | H | I | J | | | | | | | | 50016.073 |
| 197 | | | | | | | | H | I | J | K | | | | | | | 48880.945 |
| 186 | | | | | | | G | H | I | J | K | | | | | | | 48571.569 |
| 198 | | | | | | | | | I | J | K | | | | | | | 47632.468 |
| 184 | | | | | | | | | | J | K | | | | | | | 47403.703 |
| 202 | | | | | | | | | I | J | K | L | M | | | | | 42489.502 |
| 206 | | | | | | | | | | | K | L | | | | | | 41414.339 |
| 207 | | | | | | | | | I | J | K | L | M | N | O | | | 37711.033 |
| 205 | | | | | | | | | | | | L | M | N | | | | 36231.910 |
| 188 | | | | | | | | | | | | L | M | N | O | | | 33797.902 |
| 174 | | | | | | | | | | | | L | M | N | O | | | 33678.710 |
| 193 | | | | | | | | | | | | | M | N | O | | | 32221.725 |
| 181 | | | | | | | | | | | | L | M | N | O | | | 31637.020 |
| 192 | | | | | | | | | | | | | M | N | O | | | 31272.803 |
| 189 | | | | | | | | | | | | L | M | N | O | P | | 30153.093 |
| 203 | | | | | | | | | | | | | | | O | P | | 27104.339 |
| 208 | | | | | | | | | | | | | | N | O | P | | 26871.754 |
| 744 | | | | | | | | | | | | | | | | | Q | 12935.130 |
| 204 | | | | | | | | | | | | | | | | P | Q | 11925.812 |
| 199 | | | | | | | | | | | | | | | | | Q | 11172.365 |

**Table 11.** Results of a Student's *t*-test on aspect ratio from male flies both heterozygous and homozygous for a wild third chromosome. Levels not connected by the same letter are significantly different.

| Level | | Least Sq Mean |
|---|---|---|
| 199 | A | 2.2504614 |
| 192 | A B | 2.0840595 |
| 744 | B | 2.0528384 |
| 193 | B | 2.0404004 |
| 174 | B C | 1.9574636 |
| 205 | C | 1.9185688 |
| 181 | C D E | 1.8782356 |
| 206 | C D | 1.8762821 |
| 189 | C D E F G | 1.7995416 |
| 204 | B C D E F G H I J K L | 1.7894737 |
| 188 | D E F | 1.7391819 |
| 208 | E F G H I | 1.7022257 |
| 184 | F G H | 1.6946229 |
| 197 | F G H I | 1.6937741 |
| 176 | F G H I J | 1.6753813 |
| 201 | F G H I J K | 1.6567411 |
| 194 | F G H I J K | 1.6511373 |
| 203 | F G H I J K L | 1.6473466 |
| 183 | F G H I J K L | 1.6432562 |
| 190 | F G H I J K L | 1.6269901 |
| 185 | F G H I J K L | 1.6261891 |
| 195 | F G H I J K L | 1.6179691 |
| 198 | F G H I J K L | 1.6149862 |
| 445 | F G H I J K L M | 1.6121242 |
| 207 | D E F G H I J K L M | 1.6059114 |
| 209 | G H I J K L | 1.6057810 |
| 202 | F G H I J K L M | 1.5991624 |
| 177 | H I J K L M | 1.5829307 |
| 175 | H I J K L M | 1.5675666 |
| 745 | I J K L M | 1.5659596 |
| 200 | I J K L M | 1.5535688 |
| 186 | H I J K L M | 1.5449688 |
| 196 | K L M | 1.5330681 |
| 180 | J K L M | 1.5178910 |
| 179 | L M | 1.5154808 |
| 182 | M | 1.4747674 |
| 187 | M | 1.4619501 |

**Figures**

**Figure 1.**

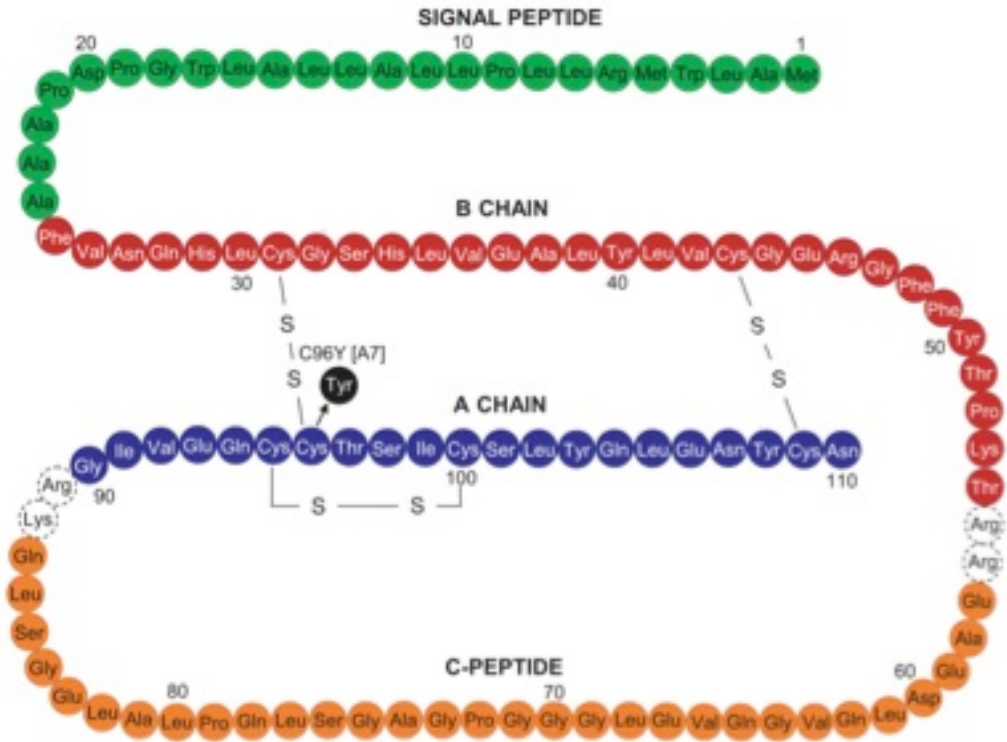

**Figure 2.**

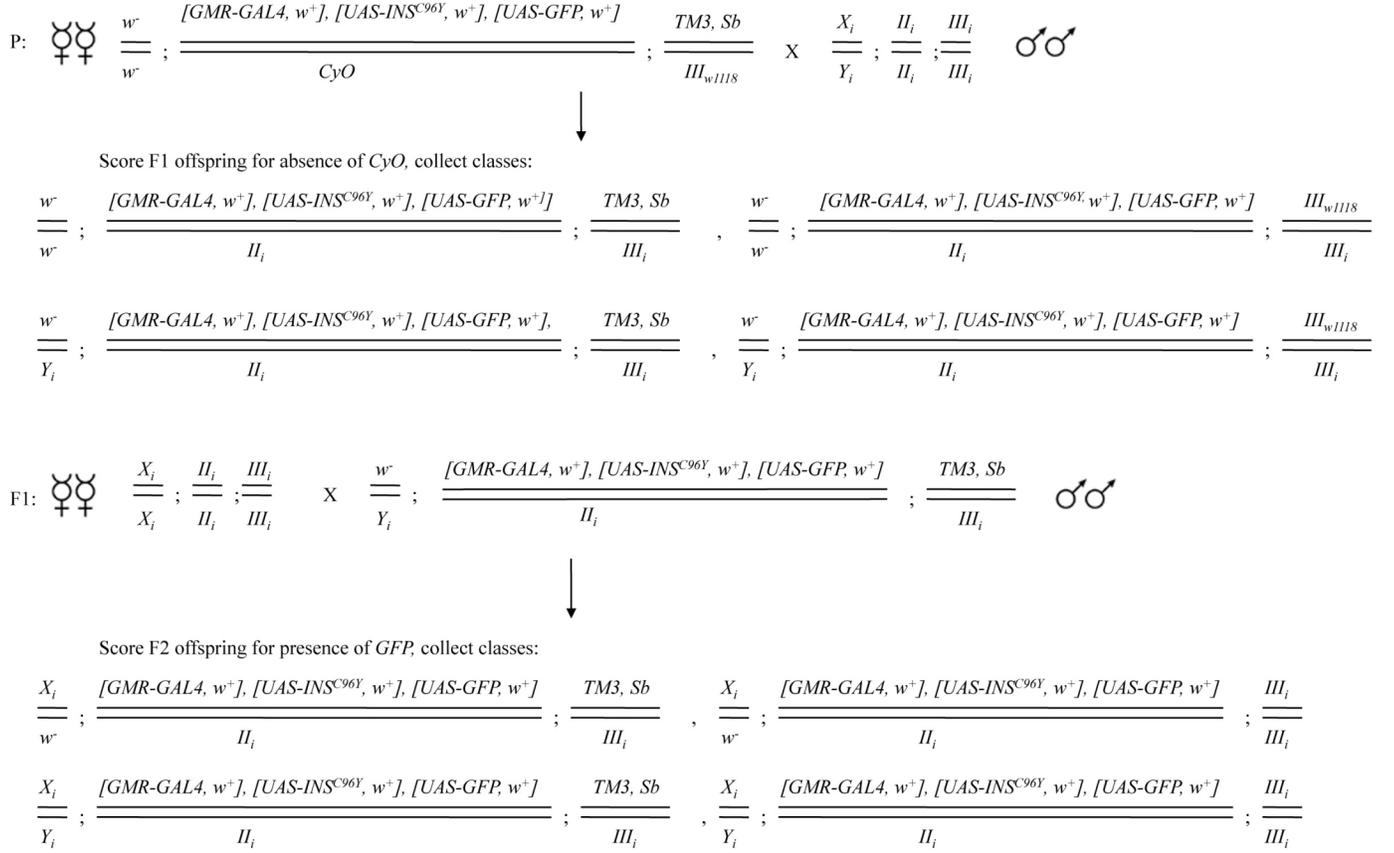

**Figure 3.**

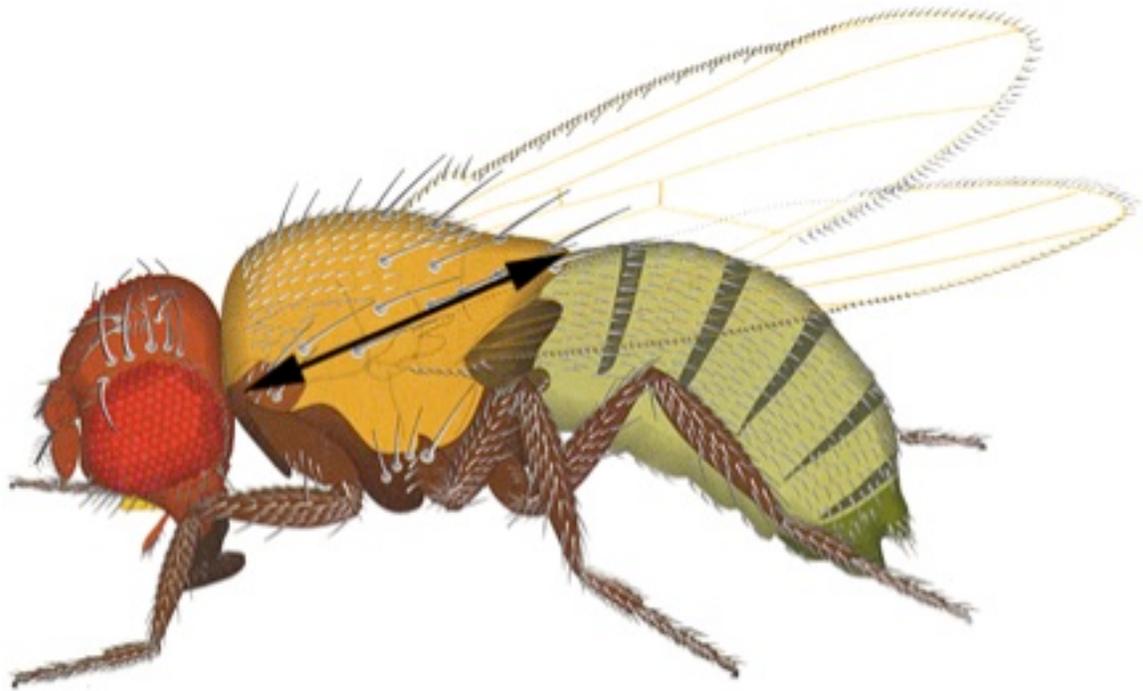

**Figure 4.**

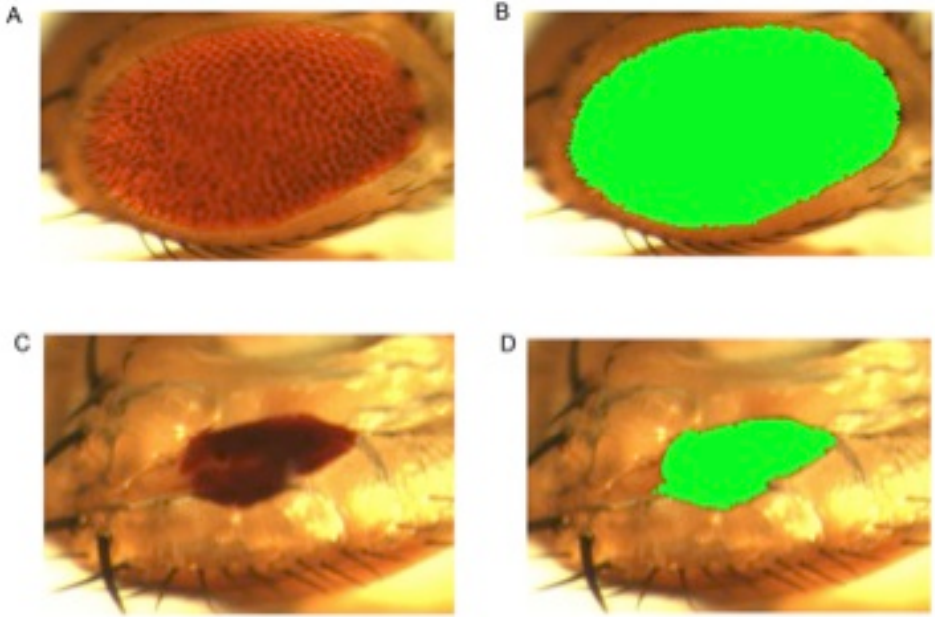

**Figure 5.**

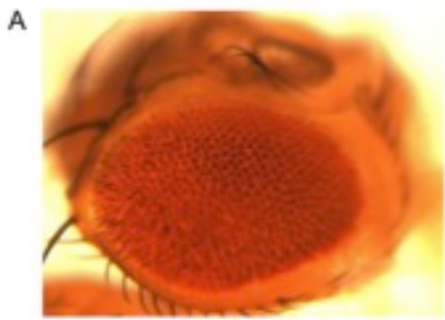

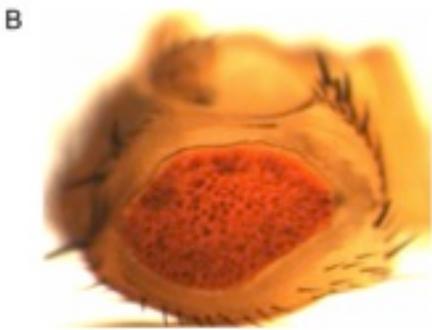

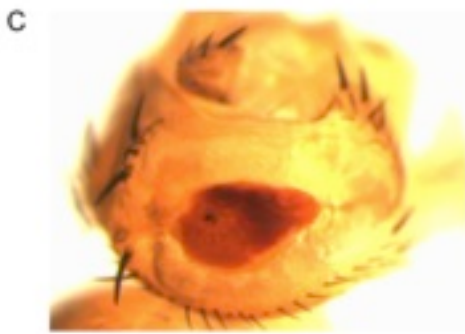

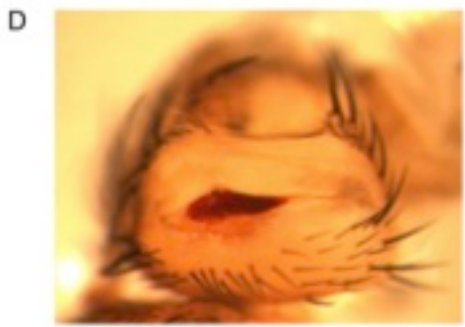

**Figure 6.**

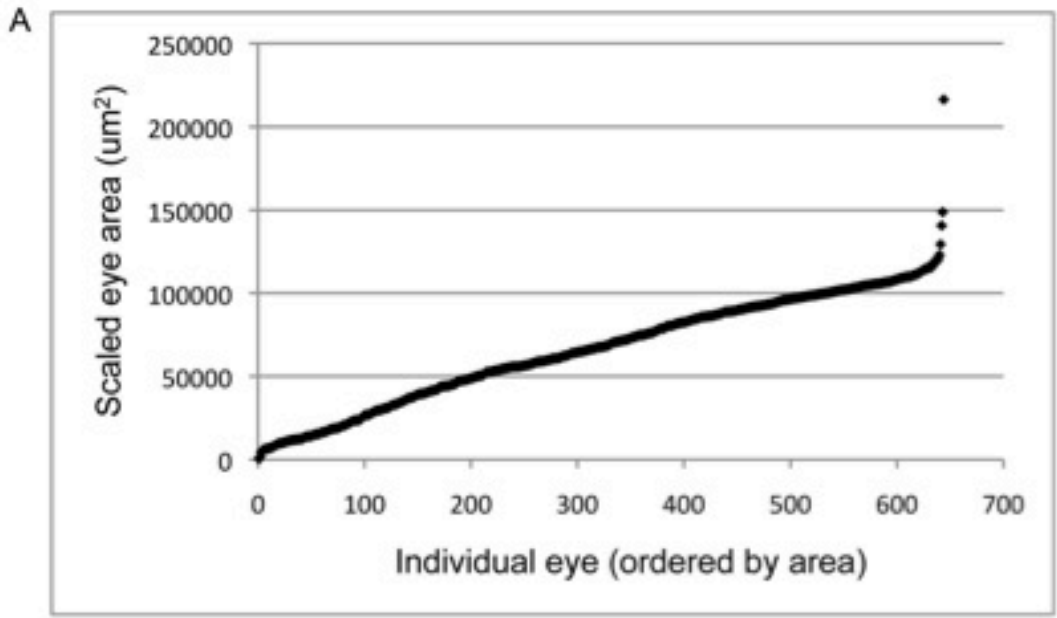

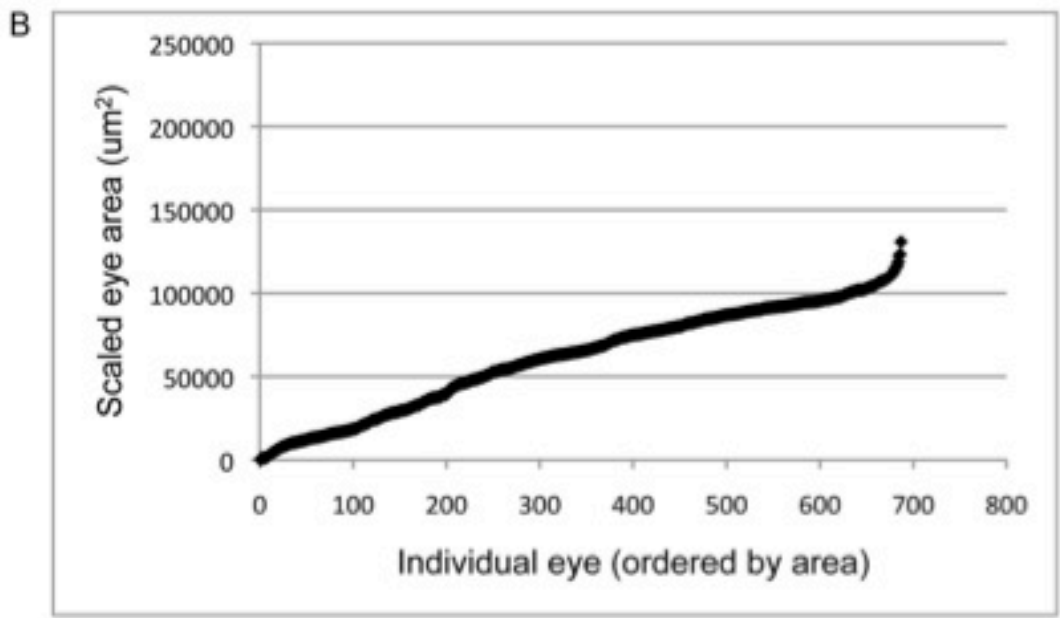

**Figure 7.**

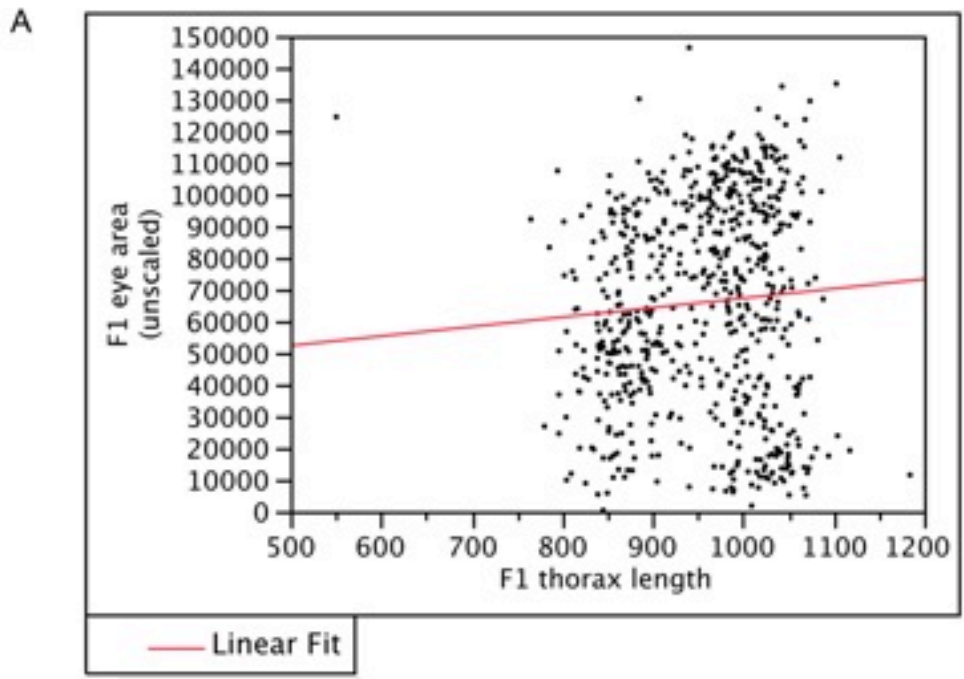

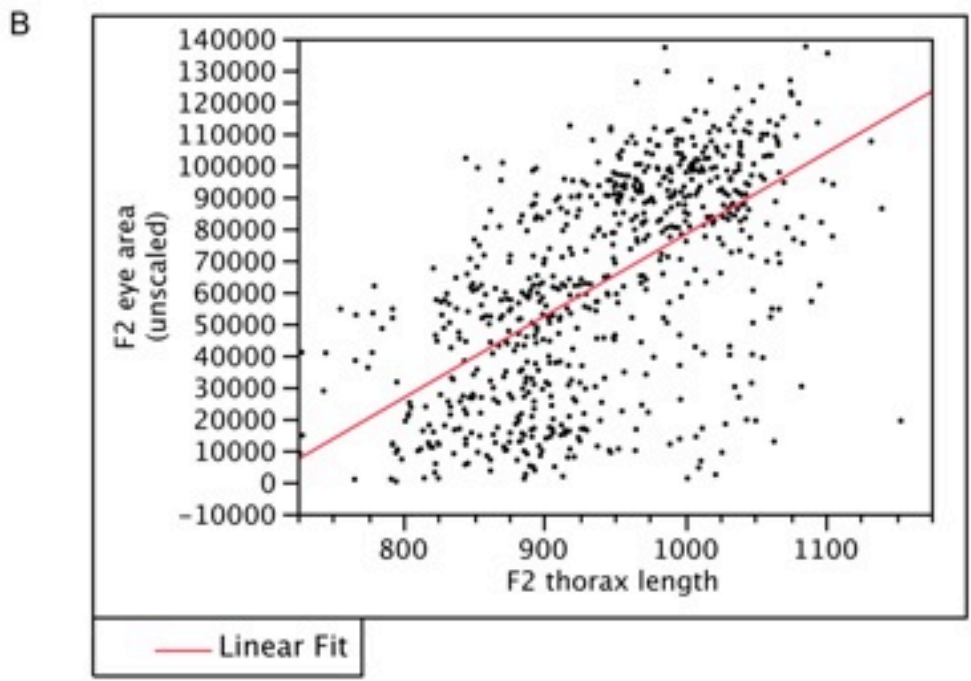

**Figure 8.**

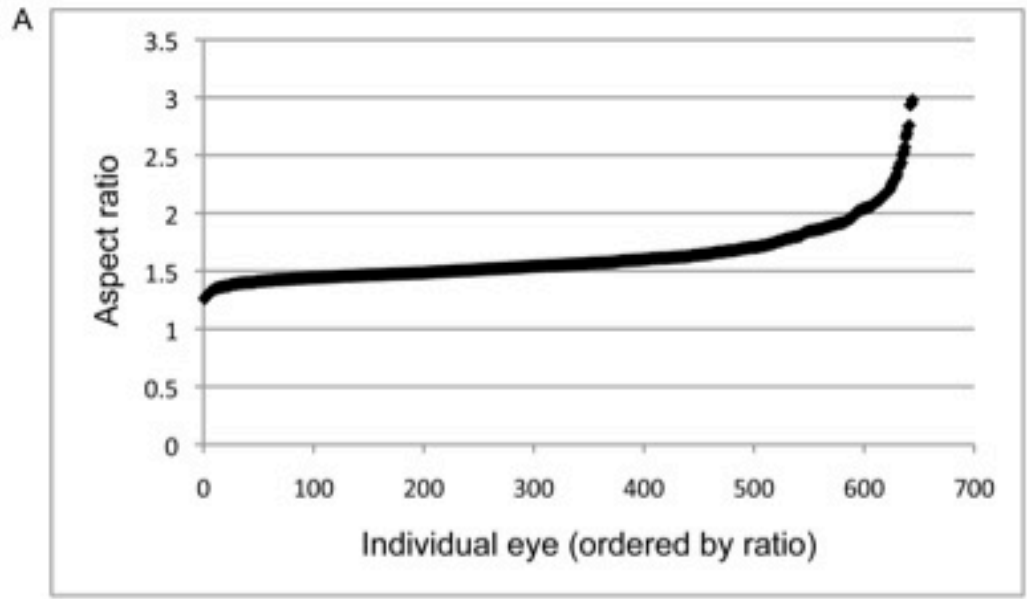

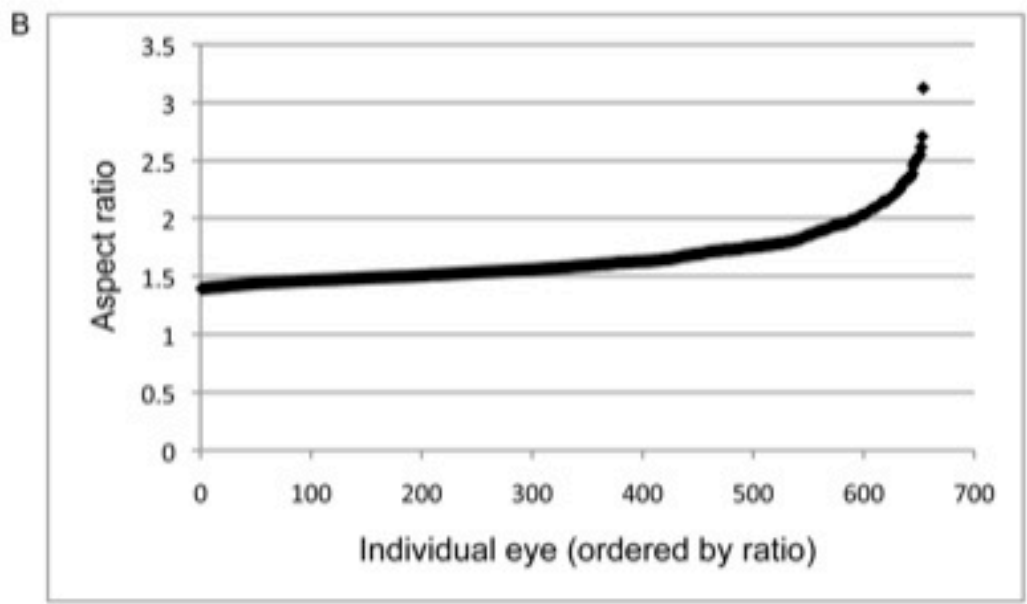

**Figure 9.**

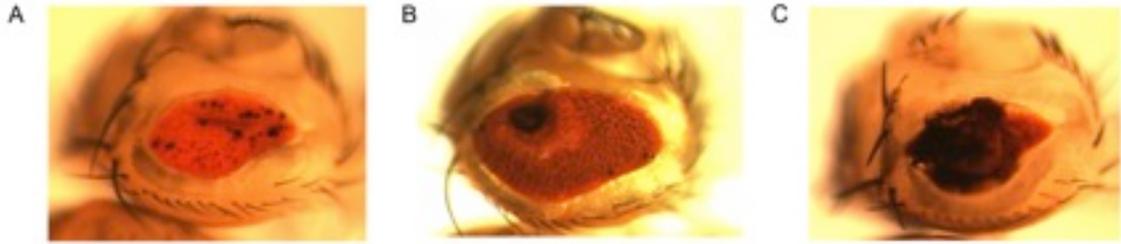

**Figure 10.**

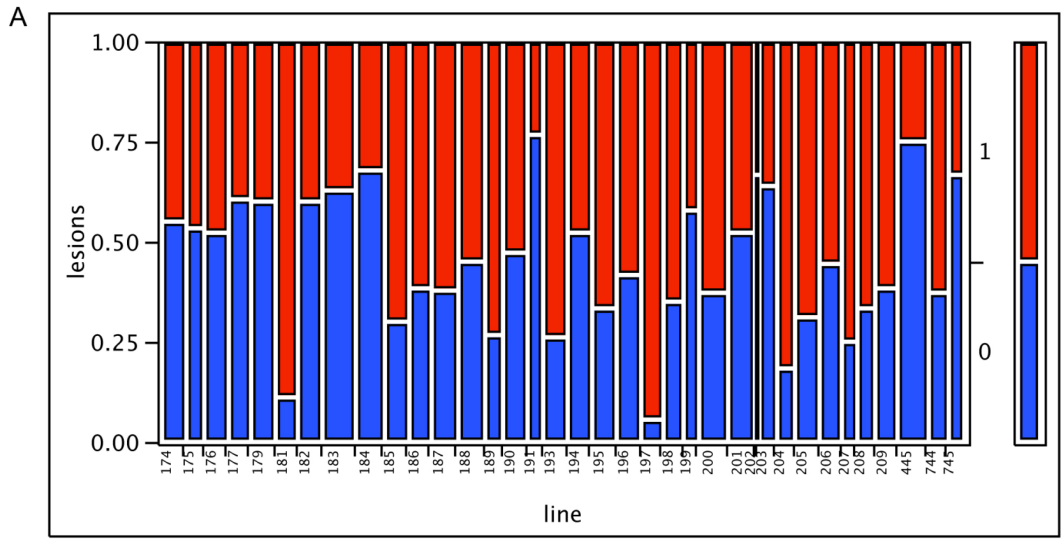

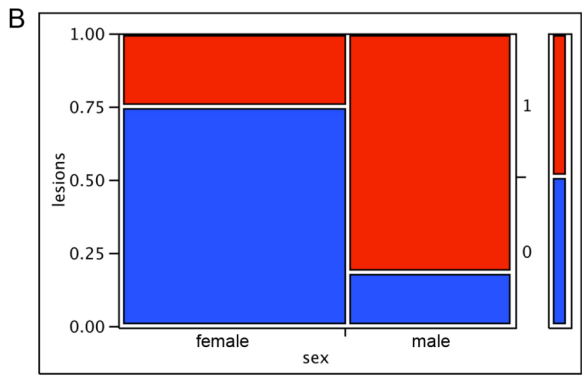

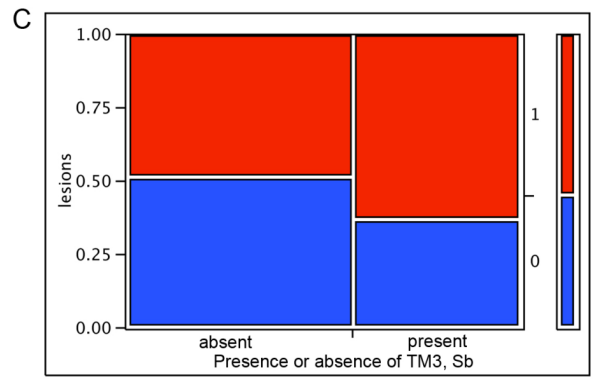

**Figure 11.**

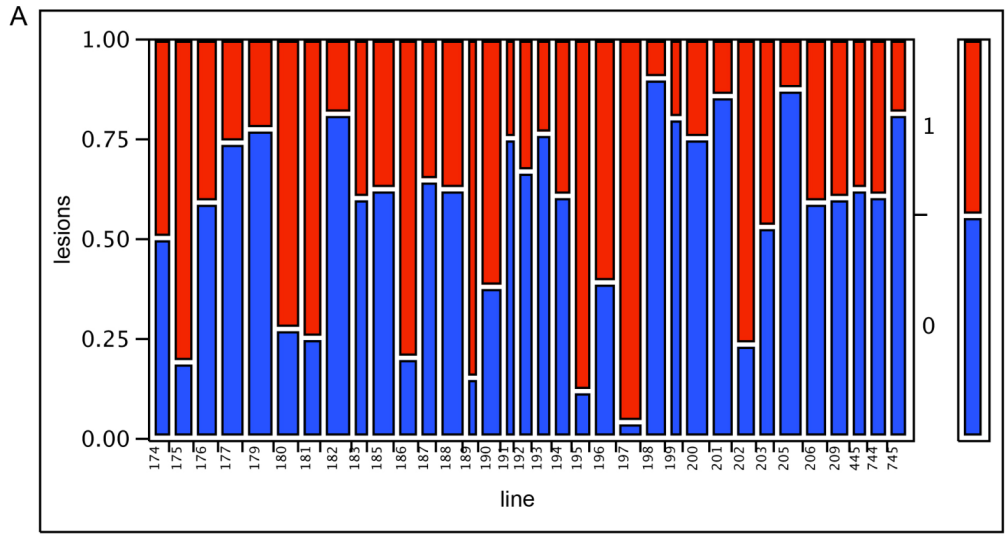

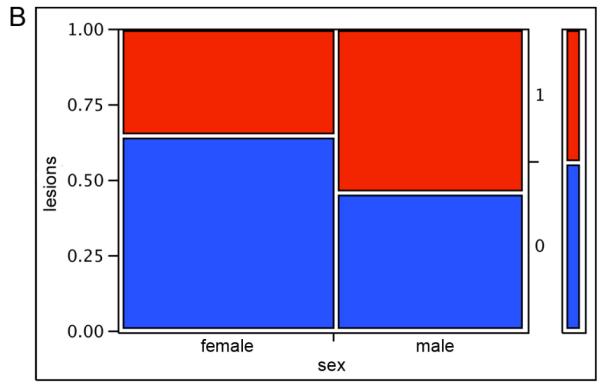
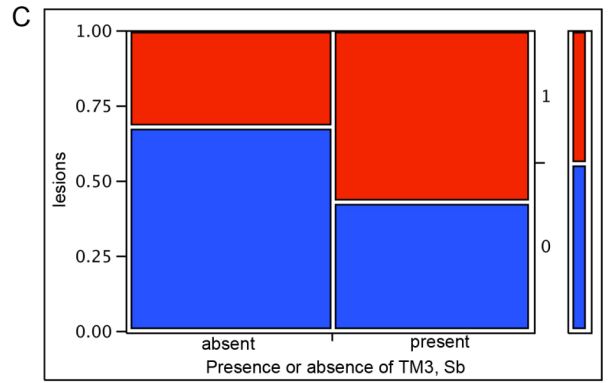

**Figure 12.**

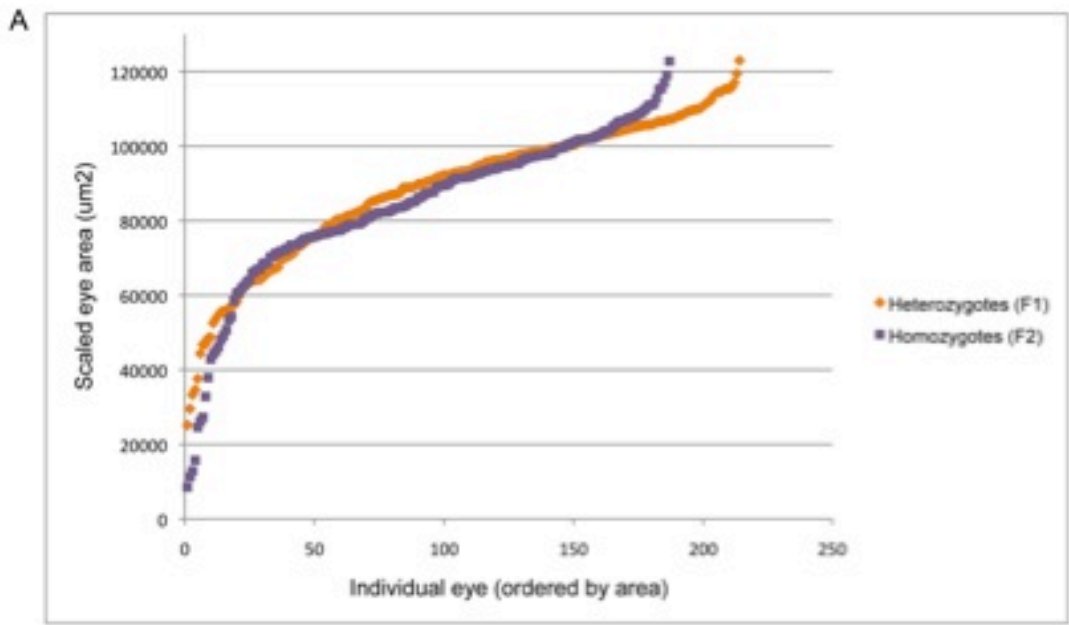

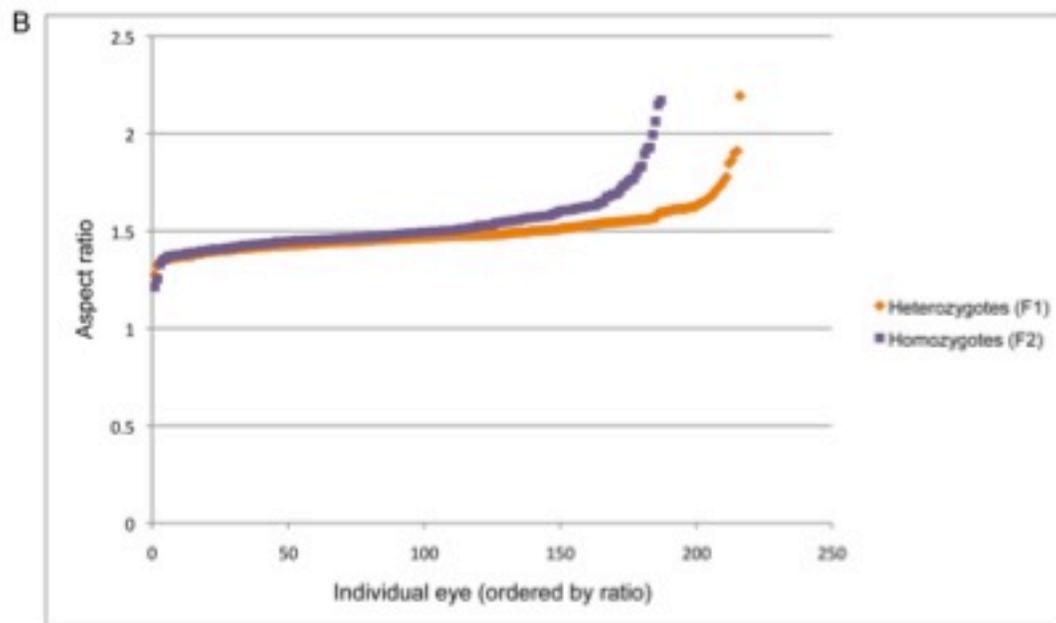

**Figure 13.**

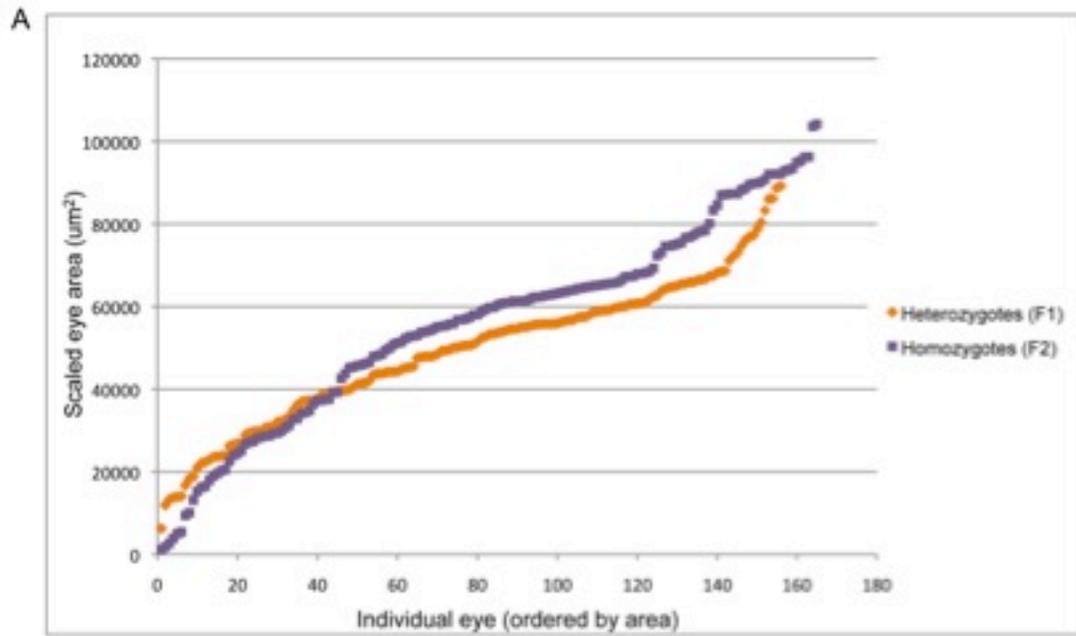

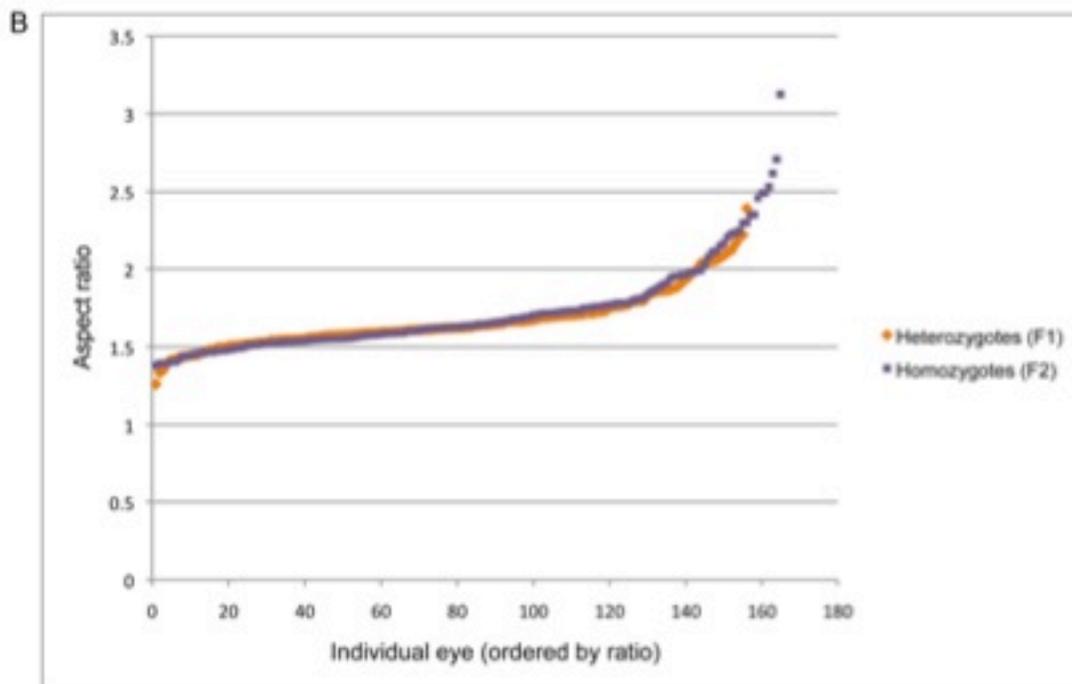